\documentclass[a4paper,11pt]{article}

\usepackage{amssymb,amsmath,amsfonts}
\usepackage{float}
\usepackage{relsize}
\usepackage{graphicx}
\usepackage{tikz}
\usetikzlibrary{patterns}
\usetikzlibrary{shapes, arrows, shadows}
\usepackage{upgreek}

\usepackage{pgfplots}

\pgfplotsset{compat=newest}

\pgfdeclarepatternformonly[\LineSpace]{my north east lines}{\pgfqpoint{-1pt}{-1pt}}{\pgfqpoint{\LineSpace}{\LineSpace}}{\pgfqpoint{\LineSpace}{\LineSpace}}%
{
    \pgfsetlinewidth{0.4pt}
    \pgfpathmoveto{\pgfqpoint{0pt}{0pt}}
    \pgfpathlineto{\pgfqpoint{\LineSpace + 0.1pt}{\LineSpace + 0.1pt}}
    \pgfusepath{stroke}
}

\newdimen\LineSpace
\tikzset{
    line space/.code={\LineSpace=#1},
    line space=3pt
}

\def\be{\begin{equation}}
\def\ee{\end{equation}}
\def\bea{\begin{eqnarray}}
\def\eea{\end{eqnarray}}

\begin{document}

\begin{titlepage}
\date{\today}       \hfill

\begin{center}

\vskip .5in

{\LARGE \bf   Open topological defects and boundary RG flows }\\
\vspace{5mm}

\today
 
\vskip .250in

\vskip .5in
{\large Anatoly Konechny} 

\vskip 0.5cm
{\it Department of Mathematics,  Heriot-Watt University\\
Edinburgh EH14 4AS, United Kingdom\\[10pt]
and \\[10pt]
Maxwell Institute for Mathematical Sciences\\
Edinburgh, United Kingdom\\[10pt]
}
E-mail: A.Konechny@hw.ac.uk
\end{center}

\vskip .5in
\begin{abstract} \large
In the context of   two-dimensional rational conformal field theories we consider topological junctions of 
topological defect lines with boundary conditions. We refer to such junctions as open topological defects. 
For  a relevant boundary operator on a conformal boundary 
condition we consider a commutation relation with an open defect obtained by passing the junction point through 
the boundary operator. We show that when there is an open defect that commutes or anti-commutes 
with the boundary operator there are interesting implications for the boundary RG flows triggered by this operator. 
The end points of the flow must satisfy certain constraints which, in essence, require the end points to admit junctions 
with the same open defects.  Furthermore, the open defects in the infrared must generate a subring under fusion that is isomorphic to the analogous subring of the original boundary condition. We illustrate these constraints by a number of explicit examples in Virasoro minimal models. 

\end{abstract}

\end{titlepage}

\renewcommand{\thepage}{\arabic{page}}
\setcounter{page}{1}
\large 

\section{Introduction }
\renewcommand{\theequation}{\arabic{section}.\arabic{equation}}
In this paper we consider Euclidean two-dimensional quantum field theories on a half-plane which are described by 
a unitary conformal field theory (CFT) in the bulk, and on the boundary by a perturbed conformal boundary condition. 
We will assume that the CFT at hand is rational and thus possesses some non-trivial topological defect lines. 
Such defects were  first considered in \cite{PZ} and then more extensively in the context of general rational CFTs in \cite{FFRS2}. They   
describe symmetries and dualities of the critical system described by the given CFT \cite{FFRS}, \cite{FFRS2}. 
Topological defect lines can be moved around and, if they do not pass though any other observables any correlation function is independent of their position. When they pass through a local bulk operator, generically we obtain 
a collection of defect segments attached to the main defect and ending on a disorder field located at the insertion.
In certain special situations the additional defect segments may be absent and passing the defect through results in 
 an operator with the same 
Virasoro representation labels but multiplied by some factor. In the simplest situation the original insertion remains intact and 
we can say that the defect commutes with this bulk insertion. 
As shown in \cite{5C}, if we have some defects that commute with 
a bulk relevant operator then there are interesting consequences for the bulk renormalisation group (RG) flows triggered by this operator. 
The fusion algebra of such commuting defects between themselves must be robust under the fusion and this places 
 constraints on the end points of the flows (triggered by the same operator with positive or negative coupling) particularly 
 when the flows are massive and the end points may be described by non-trivial topological theories.

 For a CFT on a half plane with a conformal boundary condition, if there is a non-trivial 
 boundary relevant operator, we can perturb the boundary condition by this operator triggering a boundary RG flow. 
 Unlike bulk flows boundary RG flows always end up in a non-trivial conformal boundary condition that at least has 
 the Virasoro identity tower in the boundary spectrum. In the presence of topological defect lines in the bulk CFT we 
 can fuse them with any conformal boundary condition to obtain a new conformal boundary condition which may in general be 
 a superposition of elementary boundary conditions. 
 Based on this construction, an interesting interplay between boundary RG flows and topological defect lines was discussed in \cite{GW}. The following general theorem was proved in that paper: given a boundary RG flow from a maximally symmetric 
 conformal boundary condition with label $a$ that ends in a maximally symmetric   conformal boundary condition with label $b$, 
 for any  topological defect $d$ there is an RG flow from $d\times a$ to $d\times b$ where the cross stands for fusion. 
 By maximally symmetric we mean here a boundary condition preserving  the complete chiral algebra of our rational CFT.
 We will refer to this result as Graham-Watts theorem in the rest of the paper. 
  The perturbing field for the new flow must have the same Virasoro   representation properties (and scaling dimension in particular) 
 as the perturbing  field in the original flow. As the new starting point may be a direct sum of elimentary boundary conditions 
 there may be many such fields. The precise form of the perturbing field of the new flow has been worked out in \cite{GW} for 
 the case of $a$ being elementary and for the general situation it was worked out in \cite{S}. For diagonal modular invariants 
 the action of a defect on boundary fields can be also obtained from the action on chiral defect  fields which was worked out in 
 \cite{Ingo_pertdef}.

 If we know the end-point for a particular boundary 
 flow, using Graham-Watts theorem  we can find the end-points for other flows obtained via fusion. Thus, in \cite{GW}, using the results 
 of \cite{LSS}, an extension of  perturbative flows triggered by boundary $\psi_{1,3}$ fields in minimal models \cite{RRS} to all Cardy boundary conditions was obtained. It is interesting to note that the $g$-factors change under fusion according to 
 \be \label{g-fusion}
 \sum_{i\in d\times a} g_{i} = g_{a}\left( \frac{g_{d}}{g_{1}} \right)
 \ee 
 where $g_{1}$ is the $g$-factor associated with the Cardy boundary condition that has only the identity tower in its spectrum. 
 It is not hard to show that in unitary rational CFTs $g_{1}$ is the smallest possible $g$-factor (see e.g. \cite{KF_IR}). Thus, 
  (\ref{g-fusion}) implies that fusion with non-trivial topological defect always increases the $g$-factor. A useful strategy in 
  applying the Graham-Watts theorem may then be to start with a UV boundary condition with a small value of the $g$-factor, use 
the $g$-theorem \cite{gThm1}, \cite{gThm2} and symmetries to constrain the end point as much as possible then use fusion to obtain
possible end points for flows that start with larger values of $g$.  

In this paper we look at a different usage of topological defects for constraining boundary flows that not merely relates 
two different flows but directly constrains the possible end points for a given flow. We consider   topological junctions 
of topological defects with a conformal boundary condition. This means that not only the  part of the defect line that extends into the bulk but   the junction point as well can be moved along the boundary, not changing any correlation functions as long as no boundary insertions are encountered.   Such junctions and their properties were considered at length in \cite{S} and we will use the results of that paper extensively. Following \cite{S} we call a topological defect attached to a conformal boundary via a topological junction an open topological defect. When we move such a defect along the boundary with an insertion of a boundary operator present, passing 
the open defect through the insertion typically results in a configuration with the original insertion replaced by 
several boundary condition changing fields and new 
boundary conditions  between the insertions and the open defects. But sometimes, for certain defects and boundary fields,  
no additional fields or boundary conditions arise, the open defect just passes through. In the operator language the defect 
and the boundary  operator commute. In such cases, which are similar to the bulk case considered in \cite{5C}, 
we can argue that the end point of the boundary flow must admit a topological junction with the same defect. 
Moreover, the ring obtained by fusion of such open defects between themselves must be isomorphic to some subring 
in the infrared boundary condition. This potentially can lead to additional constraints on the end points of RG flows as in the boundary case the fusion 
ring for open defects in general  depends on the boundary condition  \cite{S}. Even if the bulk labels are the same the fusion rules 
may be different.  We illustrate this on an explicit example in section \ref{penta_sec}.

Another interesting case is when an open defect just multiplies the operator by minus one when passing through it, or 
in other words when the open defect anti-commutes with the boundary operator. 
This situation demands that there must be a topological junction with the same defect and the two boundary conditions 
describing the infrared end points for the two signs of the perturbation. The fusion rules again must be robust (up to isomorphism) 
and persist into the infrared fixed points. If both commuting and anti-commuting open defects are present they form a 
${\mathbb Z}_2$-graded subring under fusion. 

The main goal of this paper is to point to the existence of such constraints on boundary RG flows, to explain how to 
look for commuting and anti-commuting open defects  and to illustrate the resulting constraints on concrete examples. 
To this end we choose to restrict our constructions to Virasoro minimal models with diagonal modular invariant. 
Moreover, our main examples of boundary flows will be the flows triggered by boundary $\psi_{1,3}$ operators. These flows 
are integrable and the end points of the flows are known. This allows us to check that the constraints we derive from open defects 
are satisfied. In addition we consider two flows: one triggered by a boundary $\psi_{2,1}$ operator in the Tetracritical Ising model and another triggered by a $\psi_{1,2}$ operator in the pentacritical Ising model. These flows are 
believed to be integrable but, to the best of our knowledge, have not been investigated before. 
We derive a number of analytic constraints on the possible end points in these flows. 

The main body of the paper is organised as follows. In section \ref{sec:defects} we discuss  generalities  about topological 
defects and their junctions with boundary conditions. We fix our normalisation conventions and derive a commutation 
relation for an open defect and a boundary operator. In section \ref{general_sec} we discuss the constraints on RG flows 
arising from open defects commuting or anti-commuting with the perturbing operator. In section \ref{sec:examples} we work 
out explicit examples in the tetracritical and the pentacritical Ising models. In section \ref{sums_sec} we discuss some specifics for 
flows triggered by boundary condition changing operators. For such flows there may be special linear combinations of 
different open defects (with the same Virasoro labels) that commute with the perturbing field. We give some explicit examples of this. 
Section \ref{conc_sec} contains some concluding remarks.
The appendix contains some useful relations between the diagonal  minimal model fusion matrices.

\section{Open topological defects} \label{sec:defects}
\setcounter{equation}{0}
Throughout the paper, except for section \ref{general_sec},  we restrict ourselves to the case of unitary Virasoro minimal models with diagonal 
modular invariant. 
For the minimal models both topological defects 
\cite{PZ} and the elementary conformal boundary conditions \cite{Cardy} are labeled by the same pairs of integers from the Kac table as the chiral  operators. 
In this section we will just use the letters: $a,b,c,\dots$ for such labels.  Boundary fields linking a boundary condition $a$ on the left with a boundary condition $b$ on the right are built on Virasoro representations $i\in a\times b$. We denote such fields 
 as $\psi_{i}^{[a,b]}$. On the diagrams below we will  omit the upper indices of boundary operators as those can be read off from 
 the boundaries. 
 
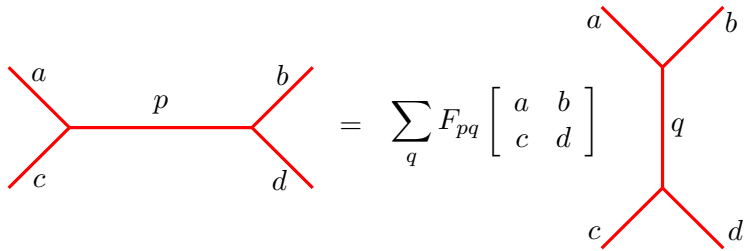
\begin{figure}[H]
\begin{center}
\begin{tikzpicture}[scale=2]
\draw[very thick, red] (-0.6,0)--(0.6,0);
\draw[very thick, red] (-0.6,0)--(-1,0.4);
\draw[very thick, red] (-0.6,0)--(-1,-0.4);
\draw[very thick, red] (0.6,0)--(1,0.4);
\draw[very thick, red] (0.6,0)--(1,-0.4);
\draw (0,0.15) node {$p$};
\draw (-0.8,0.35) node {$a$};
\draw (0.8,0.35) node {$b$};
\draw (-0.8,-0.35) node {$c$};
\draw (0.78,-0.35) node {$d$};
\draw (1.25,0) node {$=$};
\draw (2.2,0) node  {$\mathlarger{\sum\limits_{q}}\, F_{pq}\left[ \begin{array}{cc} a&b\\
c& d
\end{array} \right]$};
\draw[very thick, red] (3.3,-0.4)--(3.3,0.4);
\draw[very thick, red] (3.3,0.4)--(2.9,0.8);
\draw[very thick, red] (3.3,0.4)--(3.7,0.8);
\draw[very thick, red] (3.3,-0.4)--(2.9,-0.8);
\draw[very thick, red] (3.3,-0.4)--(3.7,-0.8);
\draw (3.4,0) node {$q$};
\draw (2.85,0.7) node {$a$};
\draw (3.75,0.7) node {$b$};
\draw (2.85,-0.7) node {$c$};
\draw (3.78,-0.68) node {$d$};
 \end{tikzpicture}
\end{center}
\caption{Elementary move in a defect network}
\label{el_move}
\end{figure}

 Three elementary topological defects labelled by $a$, $b$, $c$  can be joined together if that is permitted by the fusion, that is if 
$a\in b\times c$. Defect networks can be simplified via a sequence of elementary moves. The latter equates two networks 
 as depicted on Figure \ref{el_move}. This was shown in \cite{Ingo_nonloc} using the topological field theory approach of \cite{FFRS2}.
 
As emphasised in \cite{S} one does not have to choose the $F$-matrices appearing on Figure \ref{el_move} 
to be the same as the conformal block $F$-matrices. The latter are fixed if we canonically normalise the conformal 
blocks. To do concrete calculations we are going to use the conformal 
block $F$-matrices calculated in \cite{Runkel}, \cite{Runkel_PhD}, so we are going to assume that the defect junctions are normalised in such a way  that the defect $F$-marices are those of the conformal blocks.  We also assume that the identity defect can be attached at any point and 
can be moved freely without changing anything.

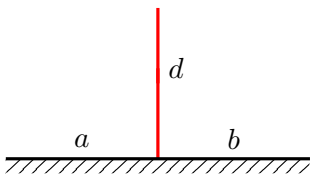
\begin{figure}[H]
\begin{center}
\begin{tikzpicture}[>=latex,scale=2]
\draw [white,pattern=my north east lines,  line space=5pt, pattern color=black] (-1,0) rectangle (1,-0.1) ;
\draw[very thick, red] (0,0)--(0,0.6);
\draw[very thick, red] (0,0.5)--(0,1);
\draw (-0.5,0.12) node {$a$};
\draw (0.5,0.13) node {$b$};
\draw (0.12,0.6) node {$d$};
\draw[very thick] (-1,0)--(1,0);

\end{tikzpicture}
\end{center}
\caption{A defect junction with a conformal boundary}
\label{junction_pic}
\end{figure}

We are further interested in topological defects that can end topologically on a given conformal boundary condition. 
This means that the ending should behave as a local operator of dimension zero. It is not hard to see that 
for an elementary boundary condition with a label $a$ and an elementary  defect with label $d$ the junction is topological 
if the fusion $d\times a$ contains $a$. To see this we can deform the defect keeping the junction pinned down so that 
the defect fuses with the  boundary on one side of the junction. The junction then looks like a boundary condition-changing 
operator between $a$ and $d\times a$. There is a dimension zero such operator if $d\times a$ contains $a$. 
Equivalently   $a\times a$ should contain $d$ and therefore the set of admissible defect labels $d$  is the same as 
the set labelling boundary operators. This observation generalises to junctions which have two different 
elementary boundary conditions on either side of the junction: $a$ and $b$. The junction is topological if there is 
a fusion vertex linking $a$, $b$ and $d$.  Such a junction is depicted on Figure  \ref{junction_pic}.

Each junction of an elementary defect and an elementary  boundary comes with a choice of coupling that can be thought of as 
a choice of normalisation of the corresponding junction field. We are going to choose the normalisation for the open defects and the boundary fields as described in \cite{S}. The conventions  of \cite{S} include additional factors for the junctions of defects with boundaries which arise from taking a defect stretched parallel to the boundary and partially fusing its right or left half with the 
boundary. To distinguish between the two types of fusion it will be convenient to orient our defects assuming that the defect outgoing 
from the boundary was fused on the left and the defect coming into the boundary was fused on the right. The orientation will 
be marked by arrows on the diagrams. Furthermore, to signify the presence of these additional factors we will  add a bullet on the junction when depicting it. The factors themselves 
are presented on  Figure \ref{bullet_factors_pic}.
 \begin{figure}[H]
\begin{center}
\begin{tikzpicture}[>=latex,scale=2]
\draw [white,pattern=my north east lines,  line space=5pt, pattern color=black] (-0.8,0) rectangle (0.8,-0.1) ;
\draw[very thick] (-0.8,0)--(0.8,0);
\draw[very thick, red,->] (0,0)--(0,0.4);
\draw[very thick, red] (0,0.3)--(0,0.6);
\draw (-0.4,0.12) node {$a$};
\draw (0.4,0.13) node {$b$};
\draw (0.12,0.4) node {$d$};

\draw (0,0) node {$\bullet$} ;
\draw (1,0) node {$=$}; 
\draw (1.8,0) node {$\sqrt{F_{1a}\left[ \begin{array}{cc} d&b\\
d& b
\end{array} \right]}$}; 
\draw [white,pattern=my north east lines,  line space=5pt, pattern color=black] (2.6,0) rectangle (4.2,-0.1) ;
\draw[very thick] (2.6,0)--(4.2,0);
\draw[very thick, red,->] (3.4,0)--(3.4,0.4);
\draw[very thick, red] (3.4,0.3)--(3.4,0.6);
\draw (3,0.12) node {$a$};
\draw (3.8,0.13) node {$b$};
\draw (3.52,0.4) node {$d$};
\end{tikzpicture}
\end{center}
\begin{center}
\begin{tikzpicture}[>=latex,scale=2]
\draw [white,pattern=my north east lines,  line space=5pt, pattern color=black] (-0.8,0) rectangle (0.8,-0.1) ;
\draw[very thick] (-0.8,0)--(0.8,0);
\draw[very thick, red] (0,0)--(0,0.6);
\draw[very thick, red,<-] (0,0.2)--(0,0.6);
\draw (-0.4,0.12) node {$a$};
\draw (0.4,0.13) node {$b$};
\draw (0.12,0.4) node {$d$};

\draw (0,0) node {$\bullet$} ;
\draw (1,0) node {$=$}; 
\draw (1.8,0) node {$\sqrt{F_{1b}\left[ \begin{array}{cc} d&a\\
d& a
\end{array} \right]}$}; 
\draw [white,pattern=my north east lines,  line space=5pt, pattern color=black] (2.6,0) rectangle (4.2,-0.1) ;
\draw[very thick] (2.6,0)--(4.2,0);
\draw[very thick, red] (3.4,0)--(3.4,0.6);
\draw[very thick, red,<-] (3.4,0.2)--(3.4,0.6);
\draw (3,0.12) node {$a$};
\draw (3.8,0.13) node {$b$};
\draw (3.52,0.4) node {$d$};
\end{tikzpicture}
\end{center}
\caption{Normalisation factors for oriented defect junctions}
\label{bullet_factors_pic}
\end{figure}


We will denote a closed elementary defect located in the bulk as ${\cal D}_{a}$ while
the open defect  corresponding to the left hand side of the first diagram on Figure \ref{bullet_factors_pic} will be denoted 
as ${\cal D}_{d}^{[a,b]}$ and, for brevity,  we will write 
${\cal D}_{d}^{[a]}$ instead of ${\cal D}_{d}^{[a,a]}$. We
 will write ${\cal D}_{d}^{[a]}(t)$  to denote an insertion of   such a defect ending at point $t$ on the boundary inside 
 correlation functions of boundary operators. Also we will  denote as  ${\cal D}_{d}^{[a]}$  the corresponding operator acting on the radial quantisation 
 states on a half plane with the  boundary condition $a$. 

With  the factors given on Figure \ref{bullet_factors_pic} two simple relations hold. Firstly,  a defect arc attached to the boundary with no insertions can be 
shrunk leaving no additional  factors. This is illustrated on Figure \ref{pic:bubble}.  
\begin{figure}[H]
\begin{center}
\begin{tikzpicture}[>=latex,scale=2]
\draw[very thick] (-1,0) -- (1,0);
\draw (-0.9,0)--(-1,-0.1);
\draw (-0.8,0)--(-0.9,-0.1);
\draw (-0.7,0)--(-0.8,-0.1);
\draw (-0.6,0)--(-0.7,-0.1);
\draw (-0.5,0)--(-0.6,-0.1);
\draw (-0.4,0)--(-0.5,-0.1);
\draw (-0.3,0)--(-0.4,-0.1);
\draw (-0.2,0)--(-0.3,-0.1);
\draw (-0.1,0)--(-0.2,-0.1);
\draw (0,0)--(-0.1,-0.1);
\draw (0.1,0)--(0,-0.1);
\draw (0.2,0)--(0.1,-0.1);
\draw (0.3,0)--(0.2,-0.1);

\draw (0.4,0)--(0.3,-0.1);
\draw (0.5,0)--(0.4,-0.1);
\draw (0.6,0)--(0.5,-0.1);
\draw (0.7,0)--(0.6,-0.1);
\draw (0.8,0)--(0.7,-0.1);
\draw (0.9,0)--(0.8,-0.1);
\draw (1,0)--(0.9,-0.1);

\draw (-0.8,0.15) node {$a$};
\draw (0,0.15) node {$b$};
\draw (0.8,0.15) node {$c$};
\draw[very thick, red] (-0.5,0) arc (180:0:0.5 );
\draw[very thick, red,->] (0.05,0.5)--(0.06,0.5);
\draw (-0.2, 0.63) node {$d$};
\draw (-0.5,0) node {$\bullet$} ;
\draw (0.5,0) node {$\bullet$} ;

\draw (1.3,0) node {$=$};
\draw (1.8,0) node {$\mathlarger{\mathlarger{\delta_{ac} } }  $};

\draw [white,pattern=my north east lines,  line space=5pt, pattern color=black] (2.2,0) rectangle (3.7,-0.1) ;
\draw (3,0.15) node {$a$};
\draw[very thick] (2.2,0)--(3.7,0);

\end{tikzpicture}
\end{center}
\caption{Shrinking an open defect bubble}
\label{pic:bubble}
\end{figure}

Secondly, when  we partially fuse a portion of a defect with the boundary we obtain a sum over elementary 
boundary conditions appearing in the fusion and two junctions with the boundary. This is illustrated on the following 
diagram
\begin{figure}[H]
\begin{center}
\begin{tikzpicture}[scale=2,>=latex]

\draw [white,pattern=my north east lines,  line space=5pt, pattern color=black] (-0.8,0) rectangle (0.8,-0.1) ;
\draw (0,0.13) node {$a$};
\draw[very thick] (-0.8,0)--(0.8,0);
\draw[ very thick, red] (-0.8,0.3) -- (0.8,0.3);
\draw[ very thick, red,->] (-0.8,0.3) -- (0.1,0.3);
\draw (0,0.45) node {$d$};
\draw (1,0) node {$=$};
\draw (1.9,0) node {$\mathlarger{\sum\limits_{a'\in d\times a} }$};

\draw [white,pattern=my north east lines,  line space=5pt, pattern color=black] (2.6,0) rectangle (4.4,-0.1) ;
\draw (3.5,0.14) node {$a'$};
\draw (2.8,0.13) node {$a$};
\draw (4.2,0.13) node {$a$};
\draw[very thick] (2.6,0)--(4.4,0);

\begin{scope}[very thick, red, every node/.style={sloped,allow upside down}]
\draw[->] (2.6,0.3) to[out=0,in=90]  (3.3,0); 
\draw[->] (3.7,0) to[out=90,in=180] (4.4,0.3);
\end{scope}
\draw (3.3,0) node  {$\bullet$} ;
\draw (3.7,0) node  {$\bullet$} ;
\end{tikzpicture}
\caption{Partial fusion of defect with a boundary}
\end{center}
\end{figure}
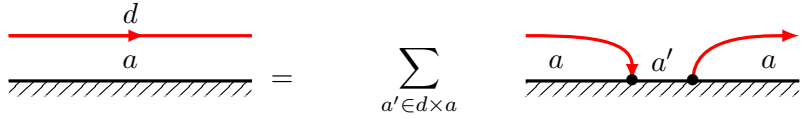

\begin{figure}[H]
\begin{center}
\begin{tikzpicture}[scale=2,>=latex]
\draw [white,pattern=my north east lines,  line space=5pt, pattern color=black] (-0.8,0) rectangle (0.8,-0.1) ;
\draw[very thick] (-0.8,0) -- (0.8,0);
\draw (-0.5,0.13) node {$a$} ; 
\draw (0.5,0.13) node {$b$} ; 
\draw (0,0) node {$\bullet$};
\draw (0,-0.23) node {$\psi_{i}$};
\draw (1,0) node {$=$};
\draw [white,pattern=my north east lines,  line space=5pt, pattern color=black] (1.2,0) rectangle (3,-0.1) ;
\draw[very thick] (1.2,0) -- (3,0);
\begin{scope}[very thick, red, every node/.style={sloped,allow upside down}]
\draw[->] (1.2,0.6) to[out=0,in=180]  (2.1,0); 
\draw[->] (2.1,0) to[out=0,in=180] (3,0.6);
\end{scope}
\draw (2.1,0) node {$\bullet$}; 
\draw (1.5,0.15) node {$1$};
\draw (2.7,0.15) node {$1$};
\draw (2.1,-0.23) node {$\psi_{i}$};
\draw (1.4,0.7) node {$a$};
\draw (2.8,0.7) node {$b$};
\draw (3.2,0) node {$=$};
\draw [white,pattern=my north east lines,  line space=5pt, pattern color=black] (3.4,0) rectangle (5.2,-0.1) ;
\draw[very thick] (3.4,0) -- (5.2,0);
\draw (4.3,0) node {$\bullet$}; 
\draw (4.3,-0.25) node {$\alpha_{i}^{ab}\psi_{i}$}; 
\draw[very thick, red] (3.4,0.6) -- (5.2,0.6);
\draw[very thick, red,->] (3.4,0.6) -- (4,0.6);
\draw[very thick, red,->] (4.7,0.6) -- (4.8,0.6);
\draw[very thick, red] (4.3,0.6) -- (4.3,0);
\draw[very thick, red,->] (4.3,0) -- (4.3,0.4);
\draw (3.7,0.15) node {$1$};
\draw (4.9,0.15) node {$1$};
\draw (4.43,0.33) node {$i$};
\draw (3.7,0.73) node {$a$};
\draw (4.9,0.75) node {$b$};
\end{tikzpicture}
\end{center}
\caption{Trading a boundary field for a defect ending field}
\label{def_net}
\end{figure}
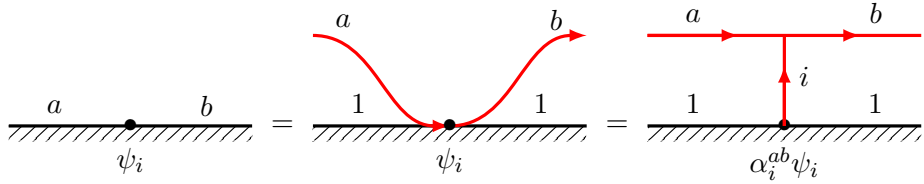

Manipulations with junctions of defects with a boundary can be lifted to 
junctions between 
topological defects by representing the boundary conditions with label $a$ as fusions between ${\cal D}_{a}$ and the 
identity boundary condition. A boundary operator with Virasoro label $i$ can be traded  for the  defect 
 labelled by $i$  ending with a defect ending field located on the identity boundary condition. 
 This is shown on Figure \ref{def_net}.
The general expression for coefficients $\alpha_{i}^{ab}$ has been calculated in \cite{S} (see equation (B.7)  of that paper). 
Once the $F$-matrices appearing in defect junctions have been fixed, these coefficients can be explicitly calculated. 
In this paper we use the conformal block $F$-matrices so in principle $\alpha_{i}^{ab}$ are fixed but at no point  in our calculations we 
need  to use  their explicit form. Using Figure \ref{def_net} we calculate, following \cite{S}, the action of an open defect on boundary operators. The latter is obtained by surrounding a boundary operator by the defect arc and shrinking the arc onto the operator. 
This can be calculated by a sequence of moves shown on Figure  \ref{action_open} where we consider the most general boundary condition changing operator. 


\begin{figure}[H]
\begin{center}
\begin{tikzpicture}[scale=2,>=latex]
\draw [white,pattern=my north east lines,  line space=5pt, pattern color=black] (-1,0) rectangle (1,-0.1) ;
\draw[very thick] (-1,0) -- (1,0);
\draw (0,0) node {$\bullet$};
\draw (0,-0.23) node {$\psi_{i}$};
\draw (-0.8,0.12) node {$a$};
\draw (-0.3,0.15) node {$a'$};
\draw (0.3,0.15) node {$a''$};
\draw (0.8,0.15) node {$\tilde a$};
\draw[very thick, red] (-0.5,0) arc (180:0:0.5 );
\draw[very thick, red,->] (0.05,0.5)--(0.06,0.5);
\draw (-0.2, 0.63) node {$d$};
\draw (-0.5,0) node {$\bullet$} ;
\draw (0.5,0) node {$\bullet$} ;

\draw (1.2,0) node {$=$};

\draw (2.6,0) node {$ \sqrt{F_{1\tilde a}\left[ \begin{array}{cc} d&a''\\
d& a''
\end{array} \right] F_{1a}\left[ \begin{array}{cc} d&a'\\
d& a'
\end{array} \right]}$};

\draw [white,pattern=my north east lines,  line space=5pt, pattern color=black] (4,0) rectangle (5.8,-0.1) ;
\draw[very thick] (4,0) -- (5.8,0);
\draw (4.4,0.15) node {$1$};
\draw (5.4,0.15) node {$1$};
\draw (4.7,-0.28) node {$\alpha_{i}^{a'a''}\psi_{i}$}; 
\draw[red, very thick] (4.9,0) -- (4.9,0.6);
\draw[red, very thick,->] (4.9,0) -- (4.9,0.4);
\draw[red, very thick,->] (4,0.6) -- (4.3,0.6);
\draw[red, very thick,->] (4.6,0.6) -- (4.8,0.6);
\draw[red, very thick,->] (5,0.6) -- (5.3,0.6);
\draw[red, very thick,->] (5.6,0.6) -- (5.7,0.6);
\draw[red, very thick] (4,0.6) -- (5.8,0.6);
\draw[very thick, red] (4.4,0.6) arc (180:0:0.5 );
\draw[very thick, red,->] (4.95,1.1)--(4.96,1.1);
\draw (5,0.34) node {$i$};
\draw (4.2,0.72) node {$a$};
\draw (5.6,0.73) node {$\tilde a$};
\draw (4.65,0.75) node {$a'$};
\draw (5.15,0.75) node {$a''$};
\draw (4.6,1.16) node {$d$};
\draw (4.9,0) node {$\bullet$};
\end{tikzpicture}
\end{center}
\begin{center}
\begin{tikzpicture}[scale=2,>=latex]
\draw (0,0) node {$=\mathlarger{\sum\limits_{q}}\, F_{a''q}\left[ \begin{array}{cc} a'&d\\
i& \tilde a
\end{array} \right] \sqrt{F_{1\tilde a}\left[ \begin{array}{cc} d&a''\\
d& a''
\end{array} \right] F_{1a}\left[ \begin{array}{cc} d&a'\\
d& a'
\end{array} \right]}$};
\draw [white,pattern=my north east lines,  line space=5pt, pattern color=black] (2.8,0) rectangle (4.6,-0.1) ;
\draw[very thick] (2.8,0) -- (4.6,0);
\draw (4.2,0.15) node {$1$};
\draw (3.2,0.15) node {$1$};
\draw (3.5,-0.28) node {$\alpha_{i}^{a'a''}\psi_{i}$};
\draw[red, very thick] (3.7,0) -- (3.7,0.6);
\draw[red,very thick,->] (3.7,0) -- (3.7,0.4);
\draw (3.8,0.34) node {$i$};
\draw[red,very thick] (2.5,0.6) -- (4.6,0.6);
\draw[red,very thick,->] (2.6,0.6) -- (2.7,0.6);
\draw[very thick, red] (2.8,0.6) arc (180:0:0.3 );
\draw[red,very thick,->] (3.15,0.9) -- (3.16,0.9);
\draw[red,very thick,->] (3.2,0.6) -- (3.3,0.6);
\draw[red,very thick,->] (3.5,0.6) -- (3.65,0.6);
\draw[red,very thick,->] (4.2,0.6) -- (4.3,0.6);
\draw (2.6,0.72) node {$a$};
\draw (4.1,0.72) node {$\tilde a$};
\draw (3.1,0.72) node {$a'$};
\draw (3.55,0.72) node {$q$};
\draw (3.1,1.04) node {$d$};
\draw (3.7,0) node {$\bullet$};
\end{tikzpicture}
\end{center}
\begin{center}
\begin{tikzpicture}[scale=2,>=latex]
\draw (0,0) node {$= F_{a''a}\left[ \begin{array}{cc} a'&d\\
i& \tilde a
\end{array} \right] \sqrt{\displaystyle \frac{F_{1\tilde a}\left[ \begin{array}{cc} d&a''\\
d& a''
\end{array} \right] }{F_{1a}\left[ \begin{array}{cc} d&a'\\
d& a'
\end{array} \right]}}$};
\draw [white,pattern=my north east lines,  line space=5pt, pattern color=black] (1.6,0) rectangle (3.4,-0.1) ;
\draw[very thick] (1.6,0) -- (3.4,0);
\draw[red, very thick] (2.5,0)--(2.5,0.6);
\draw[red, very thick,->] (2.5,0)--(2.5,0.4);
\draw (2.6,0.34) node {$i$};
\draw (2.5,0) node {$\bullet$};
\draw (2.3,-0.28) node {$\alpha_{i}^{a'a''}\psi_{i}$};
\draw (2,0.15) node {$1$};
\draw (3,0.15) node {$1$};
\draw (2,0.72) node {$a$};
\draw (3,0.75) node {$\tilde a$};
\draw[red, very thick] (1.6,0.6)--(3.4,0.6);
\draw (3.8,0) node {$=\, X_{i,a\tilde a}^{a'a''}$};
\draw [white,pattern=my north east lines,  line space=5pt, pattern color=black] (4.3,0) rectangle (5.3,-0.1) ;
\draw[very thick] (4.3,0) -- (5.3,0);
\draw (4.8,0) node {$\bullet$};
\draw (4.8,-0.25) node {$\psi_{i}$};
\draw (4.55,0.12) node {$a$};
\draw (5.05,0.12) node {$\tilde a$};
\draw[red, very thick] (1.6,0.6)--(3.4,0.6);
\draw[red, very thick,->] (2.2,0.6)--(2.3,0.6);
\draw[red, very thick,->] (3.2,0.6)--(3.3,0.6);
\end{tikzpicture}
\end{center}
\caption{Action of an open defect on a boundary field}
\label{action_open}
\end{figure}
The final factors $X_{i,a\tilde a}^{a'a''}$ that appear on Figure \ref{action_open} are 
\be
\label{A_shrink}
X_{i,a\tilde a}^{a'a''}= F_{a''a}\left[ \begin{array}{cc} d&a'\\
\tilde a& i
\end{array} \right] \sqrt{\frac{F_{1\tilde a}\left[ \begin{array}{cc} d&a''\\
d& a''
\end{array} \right]}{F_{1a}\left[ \begin{array}{cc} d&a'\\
d& a'
\end{array} \right]}} \left( \frac{\alpha_{i}^{a'a''}}{\alpha_{i}^{a\tilde a}}\right)\, .
\ee
An alternative derivation of this result can be done using the three-dimensional topological quantum field theory 
representation developed in \cite{FFFS} and \cite{FRS}.

\begin{figure}[H]
\begin{center}
\begin{tikzpicture}[scale=2,>=latex]
\draw [white,pattern=my north east lines,  line space=5pt, pattern color=black] (-1,0) rectangle (1,-0.1) ;
\draw (-0.8,0.13) node {$a$};
\draw (0,0) node {$\bullet$};
\draw (0,-0.23) node {$\psi_{i}$};
\draw[very thick] (-1,0)--(1,0);
\draw[ very thick, red] (-0.5,0) -- (-0.5,0.9);
\draw[ very thick, red,->] (-0.5,0) -- (-0.5,0.6);
\draw (-0.5,0) node {$\bullet$};
\draw (-0.2,0.13) node {$a'$};
\draw (0.8,0.13) node {$a''$};
\draw (-0.63,0.6) node {$d$};
\draw (1.2,0) node {$=$};

\draw [white,pattern=my north east lines,  line space=5pt, pattern color=black] (1.4,0) rectangle (3.4,-0.1) ;
\draw [very thick] (1.4,0)--(3.4,0);
\draw (2.2,0) node {$\bullet$};
\draw (2.2,-0.23) node {$\psi_{i}$};

\draw[very thick, red] (1.8,0) arc (192:0:0.4 );

\draw (1.8,0) node {$\bullet$};
\draw[very thick, red] (2.58,0.075) -- (3,0.075);
\draw[very thick, red] (3,0.065) -- (3,0.9);
\draw[very thick, red,->] (3,0.075) -- (3,0.6);
\draw (2.87,0.6) node {$d$};
\draw (1.6,0.13) node {$a$};
\draw (2,0.13) node {$a'$};
\draw (3.2,0.13) node {$a''$};

\draw (3.6, 0) node {$=$};
\draw (3.9,-0.08) node {$\mathlarger{\sum\limits_{\tilde a\in d\times a} }$};
\draw [white,pattern=my north east lines,  line space=5pt, pattern color=black] (4.3,0) rectangle (6,-0.1) ;
\draw [very thick] (4.3,0)--(6,0);
\draw (4.95,0) node {$\bullet$};
\draw (4.95,-0.23) node {$\psi_{i}$};
\draw[very thick, red] (4.55,0) arc (180:0:0.4 );
\draw[very thick, red,->] (5,0.4)--(5.05,0.4);
\draw (4.8,0.55) node {$d$};

\draw (4.55,0) node {$\bullet$};
\draw (4.75,0.13) node {$a'$};
\draw (5.15,0.13) node {$a''$};
\draw (5.35,0) node {$\bullet$};
\draw (5.55,0.13) node {$\tilde a$};
\draw (4.4,0.11) node {$a$};
\draw (5.95,0.13) node {$a''$};
\draw[very thick, red] (5.75,0) -- (5.75,0.9);
\draw[very thick, red,->] (5.75,0) -- (5.75,0.6);
\draw (5.75,0) node {$\bullet$};
\draw (5.62,0.6) node {$d$};
\end{tikzpicture}
\end{center}
\begin{center}
\begin{tikzpicture}[scale=2,>=latex]
\draw [white,pattern=my north east lines,  line space=5pt, pattern color=black] (-0.7,0) rectangle (1,-0.1) ;
\draw[very thick] (-0.7,0) -- (1,0);
\draw (-2.2,0) node {$=$};
\draw (-1.5,-0.05) node  {$\mathlarger{\sum\limits_{\tilde a\in d\times a} X_{i,a\tilde a}^{a'a''}}$};
\draw (0,0) node {$\bullet$};
\draw (0,-0.23) node {$\psi_{i}$};
\draw[very thick, red] (0.5,0) -- (0.5,0.9);
\draw[very thick, red,->] (0.5,0) -- (0.5,0.6);
\draw (0.37,0.6) node {$d$};
\draw (-0.3,0.11) node {$a$};
\draw (0.3,0.13) node {$\tilde a$};
\draw (0.8,0.13) node {$a''$};
\draw (0.5,0) node {$\bullet$};
\end{tikzpicture}
\end{center}
\caption{Commutator of open defect with a boundary operator}
\label{commutator_figure1}
\end{figure}
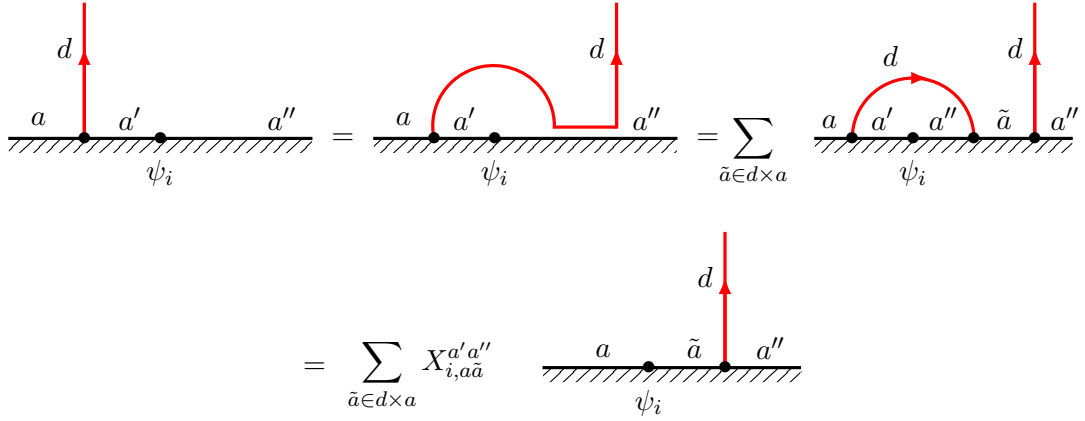

Using Figure \ref{action_open}    we can derive a commutation relation between  an open defect and an insertion of a boundary operator 
$\psi_{i}$. To that end we need to  pass the defect junction through  $\psi_{i}$ from left to right. This can be 
done by creating an arc around the insertion of $\psi_{i}$, partially fusing a portion of the defect to the 
right of the insertion and finally shrinking the arc onto the boundary field. This is depicted on Figure \ref{commutator_figure1}.

For the particular case of a boundary operator on an elementary boundary we have $a=a'=a''$ 
and the factors in the commutation relations become
\be \label{X1}
X_{i,a\tilde a}^{aa} = F_{a\tilde a}\left[ \begin{array}{cc} d&a\\
a& i
\end{array} \right] \sqrt{\frac{F_{1a}\left[ \begin{array}{cc} d&a\\
d& a
\end{array} \right]}{F_{1\tilde a}\left[ \begin{array}{cc} d&a\\
d& a
\end{array} \right]}} \left( \frac{\alpha_{i}^{aa}}{\alpha_{i}^{a\tilde a}}\right) \, .
\ee
 We see from this expression that the 
defect commutes with $\psi_{i}$ if 
\be
F_{a\tilde a}\left[ \begin{array}{cc} d&a\\
a& i
\end{array} \right] = \delta_{a\tilde a} 
\ee
and it anti-commutes if 
\be
F_{a\tilde a}\left[ \begin{array}{cc} d&a\\
a& i
\end{array} \right] = -\delta_{a\tilde a}  \, .
\ee
We can also conclude from the orthogonality relation (\ref{ort}) that these are the only interesting situations 
for RG flows originating from an elementary boundary condition, there cannot be a commutation up to a non-trivial rescaling of $\psi_{i}$. 
The latter however are possible when boundary condition changing fields are involved (see section \ref{sums_sec}).

\begin{figure}[H]
\begin{center}
\begin{tikzpicture}[scale=2,>=latex]
\draw [white,pattern=my north east lines,  line space=5pt, pattern color=black] (-1.2,0) rectangle (1.2,-0.1) ;
\draw[very thick] (-1.2,0) -- (1.2,0);
\draw[very thick, red] (-0.4,0) arc (180:0:0.4 );
\draw[very thick, red,->] (0,0.4)--(0.05,0.4);
\draw[very thick, red] (-0.9,0) arc (180:0:0.9 );
\draw[very thick, red,->] (0,0.9)--(0.05,0.9);
\draw (-0.9,0) node {$\bullet$};
\draw (-0.4,0) node {$\bullet$};
\draw (0,0) node {$\bullet$};
\draw (0.4,0) node {$\bullet$};
\draw (0.9,0) node {$\bullet$};
\draw (0,-0.23) node {$\psi_{i}$};
\draw (0.24,0.44) node {$c$};
\draw (0.45,0.92) node {$d$};
\draw (-1.1,0.13) node {$ a''$};
\draw (-0.6,0.13) node {$ a'$};
\draw (-0.2,0.11) node {$a$};
\draw (0.18,0.13) node {$ b$};
\draw (0.65,0.13) node {$ b'$};
\draw (1.15,0.13) node {$ b''$};
\draw (2.5,-0.08) node {$=\mathlarger{\sum\limits_{e\in c\times d} U_{dc}^{\{aa'a''\}[e]}U_{dc}^{\{b,b',b''\}[e]}}$};
\draw [white,pattern=my north east lines,  line space=5pt, pattern color=black] (3.8,0) rectangle (5.4,-0.1) ; 
\draw[very thick] (3.8,0) -- (5.4,0);
\draw[very thick, red] (4.15,0) arc (180:0:0.45 );
\draw[very thick, red,->] (4.6,0.45)--(4.65,0.45);
\draw (4.6,0) node {$\bullet$};
\draw (4.15,0) node {$\bullet$};
\draw (5.05,0) node {$\bullet$};
\draw (4.6,-0.23) node {$\psi_{i}$};
\draw (4,0.13) node {$ a''$};
\draw (4.4,0.11) node {$a$};
\draw (4.8,0.13) node {$b$};
\draw (5.3,0.13) node {$ b''$};
\draw (4.85,0.5) node {$e$};
\end{tikzpicture}
\end{center}
\caption{Fusion of two open defects }
\label{open_fusion}
\end{figure}
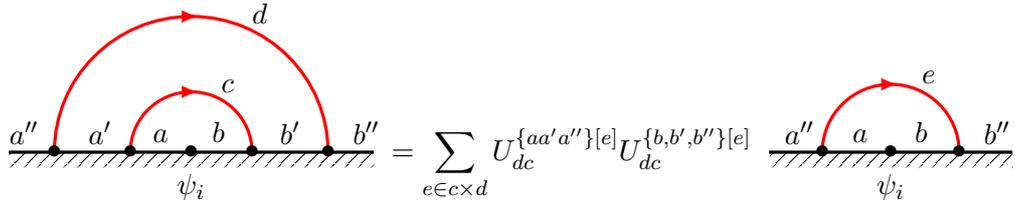

Two arcs of open defects surrounding a boundary operator can be fused into a combination of open defects 
according to Figure \ref{open_fusion}.
The coefficients on the right hand side of Figure \ref{open_fusion} were worked out in \cite{S}. They are given by 
the following combinations of the fusion matrices 
\be\label{U_coefs}
U_{dc}^{\{aa'a''\}[e]} = F_{a'e}\left[ \begin{array}{cc} d&c\\
a''& a
\end{array} \right] \sqrt{\frac{F_{1a''}\left[ \begin{array}{cc} d&a'\\
d& a'
\end{array} \right] F_{1a'}\left[ \begin{array}{cc} c&a\\
c& a
\end{array} \right]}{F_{1e}\left[ \begin{array}{cc} d&c\\
d& c
\end{array} \right]F_{1a''}\left[ \begin{array}{cc} e&a\\
e& a
\end{array} \right]}} \, .
\ee

The open defects ending on a fixed elementary conformal boundary condition $a$ are closed under fusion. 
Curiously, as noted in \cite{S},  the corresponding fusion algebra  is not given by the usual bulk fusion rule but  depends on the boundary $a$. 
In our notation we can write the deformed fusion product as 
\be \label{bfusion1}
{\cal D}_{c}^{[a]}{\cal D}_{d}^{[a]} = \sum_{e\in c\times d} { N}_{cd}^{[a]\, e} {\cal D}_{e}^{[a]}
\ee
where the coefficients $N_{cd}^{[a]\, e}$ can be obtained from Figure \ref{open_fusion} and formula (\ref{U_coefs})
by specialising to the case $a=a'=a''=b=b'=b''$. The corresponding expression can be recast into\footnote{This expression is slightly more compact than the one following from (\ref{U_coefs})  but is equivalent to it by virtue of 
$F$-matrices' identities as we show in the appendix.}
\be\label{bfusion2}
N_{cd}^{[a]\, e} =  F_{da}\left[ \begin{array}{cc} e&a\\
c& a
\end{array} \right]F_{ca}\left[ \begin{array}{cc} e&a\\
d& a
\end{array} \right]  \frac{F_{1e}\left[ \begin{array}{cc} c&d\\
c& d
\end{array} \right]}{F_{1a}\left[ \begin{array}{cc} e&a\\
e& a
\end{array} \right]}\, .
\ee
This expression is valid when the defect labeled by $e$ can end topologically on the conformal boundary  labeled by  $a$. 
It may happen that $e$ appears in the bulk fusion $c\times d$ but the defect labeled by $e$ cannot end on $a$ topologically. 
In this case $N_{cd}^{[a]\, e}$ vanishes. In general the coefficients  $N_{cd}^{[a]\, e}$ are non-negative and symmetric 
under the interchange of $c$ and $d$. The associativity of the open defect fusion was proven in \cite{S}. Among other 
general properties of  (\ref{bfusion1}), (\ref{bfusion2}) we note the following identities 
\be \label{gen_ids}
N_{1d}^{[a]\, d} =1 \, , \qquad  N^{[a] 1}_{dd} = F_{11}\left[ \begin{array}{cc} d&d\\
d& d
\end{array} \right] = \frac{S^{1}_{1}}{S_{1}^{d}}\, , \qquad \sum_{e\in c\times d} N_{cd}^{[a]\, e} =1
\ee
where $S^{i}_{j}$ is the modular $S$-matrix.

\section{Constraints on boundary RG flows}  \label{general_sec}
\setcounter{equation}{0}
Suppose now that we take an elementary conformal boundary condition labelled by $a$ and perturb it 
by a relevant operator $\psi(t)$ with a coupling $\lambda$. Let $\psi_{i}$ stand for a complete basis of local
boundary operators in the UV theory.  A renormalised  boundary correlator  in the perturbed theory can be written as  
\be \label{pert_corr1}
\langle [\psi_{i_k}](x_k) \dots [\psi_{i_1}](x_1)\rangle_{\lambda} = Z^{-1}\langle e^{-\lambda\!\! \int\limits_{-\infty}^{\infty}\!\! \psi(t) dt -S_{\rm ct}}  [\psi_{i_k}](x_k) \dots [\psi_{i_1}](x_1) \rangle 
\ee
where $S_{\rm ct}$ stands for the counterterms action,  
$[\psi_{i}]$  denote renormalised boundary operators, $\lambda$ is the renormalised coupling constant and $Z$ is the 
normalisation factor:
\be \label{Z}
Z= \langle e^{-\lambda\!\! \int\limits_{-\infty}^{\infty}\!\! \psi(t) dt -S_{\rm c}}\rangle \, .
\ee
 More explicitly we have 
\be \label{counterterms}
S_{\rm ct} = \int dt \sum_{i} M^{i}\psi_{i}(t) \, , \qquad [\psi_{i}](x) = \psi_{i}(x) + \sum_{j} M_{i}^{j}\psi_{j}(x)
\ee
where $\psi_{i}$  stand for the (bare) operators at the UV fixed point  and the terms with $M^{i}$  and $M_{i}^{j}$ 
stand for the coefficients of counterterms renormalising the action and the local operators respectively.
For brevity we are not explicitly writing  the dependence on a regulator but  assume that the regulator is  point splitting and the 
minimal subtraction scheme is employed.

Consider a correlation function at the UV fixed point in the presence of an open topological defect ${\cal D}_{d}^{[a]}$. 
It can be expressed in terms of correlators of local operators by sliding the defect to infinity along the boundary. 
Pulling the defect to the right and using the moves depicted on Figure \ref{commutator_figure1} we obtain 
\bea
&& \langle \psi_{i_k}(x_k) \dots \psi_{i_{p+1}}(x_{p+1}){\cal D}_{d}^{[a]}(s) \psi_{i_{p}}(x_{p}) \dots  \psi_{i_1}(x_1)\rangle   
\nonumber \\
&& = \langle     (\hat {\cal D}_{d}\psi_{i_k})(x_k) \dots    (\hat {\cal D}_{d}\psi_{i_{p+1}})(x_{p+1})  \psi_{i_{p}}\dots    
 \psi_{i_1}(x_1)\rangle 
 \eea 
where $x_k>\dots >x_{p+1}>s>x_{p}>\dots  >x_1$, 
\bea  \label{X_general}
(\hat {\cal D}_{d}\psi_{i_{p+1}})(x_{p+1}) &=&\sum\limits_{\tilde a \in d\times a}  \sum_{j} X_{i_{p+1},a\tilde a}^{j,aa}\psi_{j}^{[a,\tilde a]}(x_{p+1}) \, , \nonumber \\
(\hat {\cal D}_{d}\psi_{i_{l}})(x_{l}) &=&\sum\limits_{a',a''\in d\times a}  \sum_{j} X_{i_{l},a' a''}^{j,aa}\psi_{j}^{[a', a'']}(x_{l}) \, , 
\enspace l=p+1, \dots, k 
\eea
and the boundary condition $\tilde a$ is assumed to appear between  products of consecutive operators, like
$\psi^{[a',\tilde a]}_{j}(x_{q+1})\psi^{[\tilde a, a'']}_{l}(x_{q})$, where the neighbouring boundary conditions match, while 
we get zero when they do not match. The coefficients $X_{l,a' a''}^{j,aa}$ represent the shrinking bubble of the defect 
${\cal D}_{d}^{[a]}$ surrounding 
the operator $\psi_{l}^{[a,a]}$ with $j$ labelling possible degeneracies  of the Virasoro representation. For CFTs with 
minimal models type fusion\footnote{For other chiral algebras these coefficients can be computed by a sequence of moves depicted in 
Fig. \ref{action_open} but the answer will be different from (\ref{A_shrink}). It would have to take into account possible degeneracies in Virasoro representations, different fusion vertices and a charge conjugation matrix.} we have $X_{l,a' a''}^{j,aa}=\delta^{j}_{l}X_{l,a' a''}^{aa}$ where $X_{l,a' a''}^{aa}$  are given in (\ref{A_shrink}). 


Suppose now an open defect ${\cal D}_{d}^{[a]}$ commutes with $\psi$.  We note that ${\cal D}_{d}^{[a]}$ also commutes 
with the operators that appear in the counterterms action $S_{\rm ct}$.   This follows from the fact that the 
 counterterms are put in to subtract the divergences arising when several perturbing operators $\psi$ collide. Such collisions 
 can be represented by an operator product expansion   of a group of operators:
 \be \label{group_OPE}
 \psi(t_n)\psi(t_{n-1})\dots \psi(t_{1}) = \sum_{i}C^{i}(t_{1}, t_{2}, \dots , t_{n}) \psi_{i}(t_1)
 \ee 
where $C^{i}(t_{1}, t_{2}, \dots , t_{n})$ are some functions. Since ${\cal D}^{[a]}_{d}$ commutes with each operator $\psi(t_{i})$ on the left hand side, 
it commutes with each operator  $\psi_{i}(t_1)$ on the right hand side and thus with all operators appearing in $S_{\rm c.t.}$. 
This means that an insertion of ${\cal D}_{d}^{[a]}$ into a perturbed correlation function (\ref{pert_corr1}) with the 
junction located at a point $s$ can be 
moved freely inside the perturbed correlation functions as long as it does not pass through insertions of additional 
boundary operators,  that is  the correlation function 
\be
Z^{-1}\langle e^{-\lambda\!\! \int\limits_{-\infty}^{\infty}\!\! \psi(t) dt -S_{\rm ct}}  [\psi_{i_k}](x_k) \dots  {\cal D}_{d}^{[a]}(s) \dots [\psi_{i_1}](x_1) \rangle 
\ee
is independent of $s$ as long as it does not cross any of $x_1, \dots x_k$. Moreover, passing through  any of $[\psi_{i_{j}}](x_j)$ 
is given by exactly the same formulae (\ref{X_general}) as in the UV theory (with the bare 
operators $\psi_i$ replaced by $[\psi_i]$). To show this we note that the renormalised operators 
$ [\psi_{i_{j}}] $ are given by the UV operators $\psi_{i_{j}}$ plus counterterms. The latter are taken to cancel divergencies 
arising when some number of  perturbing  operators collide at the insertion point $x_j$.  Again, the counterterm operators 
are contained in the operator product expansion of a group of operators containing the operators $\psi$ 
and the UV operator $\psi_{i_{j}}$. 
Since ${\cal D}_{d}^{[a]}$ commutes with all $\psi$'s it acts on the counterterms  in the same way as it acts on $\psi_{i_{j}}$. 
This means that an insertion of ${\cal D}_{d}^{[a]}$ into a perturbed theory correlator can be traded for a linear combination 
of renormalised local correlation functions with coefficients given by those of the UV theory. We should also note that besides 
the linear combinations (\ref{X_general}) moving the defect also results in replacing the boundary condition  
$a$ between the insertions by those arising in the fusion $d\times a$ of the defect with the UV boundary condition. 
Due to the Graham-Watts theorem at the end of the flow  such segments have 
the conformal boundary condition  given by the fusion of $d$ with the  infrared BCFT.

To summarise, the above  means that ${\cal D}_{d}^{[a]}$ 
descends to a topological open defect in the perturbed theory and, consequently, at the infrared fixed point at the end of the flow. 
This places a constraint on the end point of the flow -- {\it the end point must  be given by a conformal boundary condition that admits 
a topological junction with the defect labeled by $d$}. Moreover, the action of ${\cal D}_{d}^{[a]}$ on the boundary   operators of the perturbed 
theory 
is independent of the coupling $\lambda$ -- it is given by the action in the UV BCFT. 
 Since  all open defects that commute with $\psi_i$ form a closed algebra under fusion, generated by elementary defects  ${\cal D}_{d}^{[a]}$, the same fusion rules will be valid also in the deformed theory.  Thus, in addition to admitting topological junctions with defects 
 labelled by the same $d$'s, {\it the corresponding  open defects at the end point of the RG flow must form  a subring
 under fusion that is isomorphic to that of the UV boundary condition. }
 Given that in general the fusion algebra  depends on the boundary condition  this may place some additional constraints on the IR BCFT. 
 
 
 Consider next an open defect ${\cal D}_{d}^{[a]}$ that anti-commutes with $\psi$. Let us place the corresponding junction at 
 a point $s$ on the boundary and consider a perturbation with a coupling $\lambda$ to the left of $s$ and with a coupling $-\lambda$ to the right of $s$. A deformed correlation function in such a configuration can be written as 
 \be \label{correlator+-}
Z^{-1}\langle   [\psi_{i_k}]_{-\lambda}(x_k) \dots  e^{\lambda\!\! \int\limits^{\infty}_{s}\!\! \psi(\tau) d\tau -S^{+}_{\rm ct}}{\cal D}_{d}^{[a]}(s)  
e^{-\lambda\!\! \int\limits_{-\infty}^{s}\!\! \psi(\tau) d\tau-S^{-}_{\rm ct}} \dots [\psi_{i_1}]_{\lambda}(x_1) \rangle 
\ee
where  $x_k>\dots >x_{p+1}>s>x_{p}>\dots  >x_1$, and
\be
S^{-}_{\rm ct} = \int\limits_{-\infty}^{s}\! M^{i}(\lambda) \psi_{i}(\tau)d\tau \, , \quad 
 S^{+}_{\rm ct} = \int\limits^{\infty}_{s}\! M^{i}(-\lambda) \psi_{i}(\tau)d\tau  \, , 
\ee
that is $S^{-}$ contains counterterms for the theory specified by $\lambda$ and integrated to the left of the defect and 
$S^{+}$ contains the counterterms for the theory with the coupling $-\lambda$ integrated to the right of the defect. 
Furthermore, in (\ref{correlator+-}) we have 
\be
[\psi_{i}]_{\lambda}(x) = \psi_{i}(x) + \sum_{j} M_{i}^{j}(\lambda)\psi_{j}(x)  \, , \quad 
[\psi_{i}]_{-\lambda}(x) = \psi_{i}(x) + \sum_{j} M_{i}^{j}(-\lambda)\psi_{j}(x)
\ee
so that the renormalised operators   inserted to the left of the defect are defined 
with counterterm coefficients $M_{i}^{j}(\lambda)$ corresponding to the coupling $\lambda$ while those inserted to the right have counterterms specified by $-\lambda$.

Since ${\cal D}_{d}^{[a]}$ anti-commutes with $\psi$ it commutes with  the counterterms 
that come from collisions of even numbers of $\psi$'s and anti-commutes with  those coming from collisions 
of an odd number of $\psi$'s. This means that 
\be
{\cal D}_{d}^{[a]}(\tau + \epsilon) M^{i}(\lambda) \psi_{i}(\tau) = M^{i}(-\lambda) \psi_{i}(\tau) {\cal D}_{d}^{[a]}(\tau - \epsilon) \, , \enspace \epsilon>0 \, .
\ee
Hence the correlation function in (\ref{correlator+-}) is independent of $s$ as long as $s$ does not cross any of the insertion 
points\footnote{This implies in particular that the counterterms in $S^{+}$ and $S^{-}$ are sufficient to renormalise the theory 
with the defect inserted in $s$. In particular no additional counterterms are needed to be inserted at $s$. Such additional counterterms 
would be  needed if no anti-commuting topological defect was inserted at $s$ while perturbing  with different couplings to the left and to the right 
of $s$.} $x_{1},\dots , x_{k}$. Moreover, for the same reasons as in the commuting case, when  ${\cal D}_{d}^{[a]}$ is 
being passed  from right to left through any of the insertions $[\psi_{i_j}](x_j)$  it acts on them 
via the UV theory coefficients  (\ref{X_general})  and changes the counterterms to those of the theory with the opposite coupling:
\be \label{action+-}
[\psi_{i}]_{-\lambda}(\tau) {\cal D}_{d}^{[a]}(\tau - \epsilon)  = {\cal D}_{d}^{[a]}(\tau +\epsilon) \sum\limits_{a',a''\in d\times a}  \sum_{j} X_{i_{l},a' a''}^{j,aa}[\psi_{j}^{[a',a'']}]_{\lambda}(\tau)     \, , \enspace \epsilon>0 \, 
\ee 
and the boundary condition $\tilde a$ is assumed to appear between the insertion and the new position of the defect. 
We finally comment on the  normalisation factor $Z$ in (\ref{correlator+-}). It  can be taken as in (\ref{Z}) to be 
given by the $\lambda$-deformed partition function however it  is the same if we change in   (\ref{Z}) $\lambda$ to $-\lambda$ 
as we can insert  ${\cal D}_{d}^{[a]}$ at minus infinity and move it through to plus infinity changing the sign of $\lambda$.

  Taking $\lambda$ to the  infrared fixed point, it follows from the above that we get a topological junction of the defect labeled 
 by $d$ and the two conformal boundary conditions that describe the IR endpoints of the flow in  the positive and 
 negative $\lambda$ directions. Thus, {\it for each anti-commuting defect there must exist a topological junction between the two 
 end-points of the flows in the positive and negative direction and the same bulk defect. } If we take all open defects ending on $a$ 
 that either commute or anti-commute with $\psi_i$ they form a ${\mathbb Z}_{2}$-graded algebra with respect to fusion. 
 Since the action (\ref{action+-})  is independent of $\lambda$ {\it the corresponding fusion subring  of the 
 defects at and between the infrared fixed points must be isomorphic to the one at the UV theory. }
 
 There is one other interesting constraint arising from the presence of an anti-commuting defect: {\it the $g$-factors of the 
 two infrared fixed points must be the same.} To explain why this is the case recall that the boundary entropy of the 
 perturbed boundary condition with the coupling $\lambda$ arises from the perturbed partition function on a disc 
 (see e.g. \cite{gThm1} or  \cite{gThm2}). With a point splitting regulator the value of the disc partition function remains the same if we 
 insert into it an arc with two junctions of ${\cal D}_{d}^{[a]}$ between a pair of neighbouring insertions of $\psi$. 
 We can then move one junction around the circle, anti-commuting with the insertions of $\psi$ and counterterms,  until it reaches the other junction at which point the arc can be removed. 
 This implies that the disc partition function for the coupling $\lambda$ is the same as the one with $-\lambda$ and hence the same goes 
 for their boundary entropies and the $g$-factors in the infrared fixed points.

 It should be noted that all of the above  constraints generalise in a straightforward manner to the case when the UV boundary condition is a direct sum 
 of elementary boundary conditions. 
 
Before we finish this section we would like to comment briefly on  the Hamiltonian description of the above situations. 
For simplicity we will not consider here the effects of possible divergences in the Hamiltonian formalism. 
  Consider  an infinite  strip of width $L$ with 
the  boundary condition $a$ put on both ends. Let $0 \le  \sigma \le L$ be the coordinate across the strip and $-\infty <\tau <\infty$ be the coordinate along the strip.  For  $\tau$ being Euclidean time the Hilbert space can be decomposed into Virasoro 
irreducible representations $V^{i}$  as 
\be
{\cal H}^{[a,a]} = \bigoplus\limits_{i \in a\times a}  V^{i} \, . 
\ee
Open defects that end topologically on both ends of the strip act on the states in ${\cal H}^{[a,a]}$  
by 
the action of operator ${\cal D}_{d}^{[a]}$ described in the previous section. 
Consider next perturbing the boundary condition $a$ on one or both ends of the strip by a relevant operator $\psi$.
For a perturbation on the lower end the perturbed Hamiltonian acting on ${\cal H}^{[a,a]}$
can be written as 
\be
H_{\lambda} = \frac{\pi}{L} \left[ L_{0}^{\rm UV} -\frac{c}{24} + \lambda L \psi(0,0) \right] 
\ee
where $c$ is the central charge and $L_{0}^{\rm UV}$ is the dilation operator acting on ${\cal H}^{[a,a]}$.
If an open  defect  ${\cal D}_{d}^{[a]}$ commutes with $\psi$ then  for any $\lambda$  
\be
{\cal D}_{d}^{[a]} H_{\lambda} = H_{\lambda} {\cal D}_{d}^{[a]} 
\ee
 and in particular at the IR fixed point   ${\cal D}_{d}^{[a]} $ should commute with $L_{0}$ and thus ${\cal D}_{d}^{[a]} $
 gives a symmetry of the infrared spectrum. 
 
 If  ${\cal D}_{d}^{[a]}$ anti-commutes with $\psi$ then we have 
 \be
{\cal D}_{d}^{[a]} H_{\lambda} = H_{-\lambda} {\cal D}_{d}^{[a]} \, .
\ee 
 Taking $\lambda$ to the fixed point (which is typically at infinity) we obtain 
 \be
 {\cal D}_{d}^{[a]} L_{0}^{\rm IR, 1} = L_{0}^{\rm IR, 2}  {\cal D}_{d}^{[a]} 
 \ee
 where $L_{0}^{\rm IR, 1}$ and $L_{0}^{\rm IR, 2} $ are the dilation operators for the IR endpoints corresponding to the negative and positive $\lambda$ respectively. Thus, $ {\cal D}_{d}^{[a]} $ intertwines the spectra of the two end-points. 
 
\section{Examples} \label{sec:examples}
\setcounter{equation}{0}
\subsection{ Diagonal unitary minimal models}
Here we remind the reader some basic facts about the unitary Virasoro minimal models with diagonal 
modular invariant. 
Such models are  labeled by an integer $m$ and have the central charge 
\be
c_{m} =  1 - \frac{6}{m(m+1)} \, .
\ee
The primary fields $\phi_{r,s}$ are labelled by 
two integers $1\le r\le m-1$, $1\le s\le m$ from the Kac table with the identification 
\be  \label{Kac_identification}
\phi_{r,s}\equiv \phi_{m-r,m+1-s}\, .
\ee 
 The same set of integers label the bulk defects ${\cal D}_{r,s}$ as well as the elementary conformal boundary conditions 
 which we will denote as $(r,s)$. 
 
The fusion rules are summarised in the following equation 
\be
\phi_{r,s}\times \phi_{r',s'} = \sum_{p,q}{\cal N}_{(r,s),(r',s')}^{(p,q)}\phi_{p,q} \, , \quad 
{\cal N}_{(r,s),(r',s')}^{(p,q)} = {\cal N}_{r,r'}^{p}(m) {\cal N}_{s,s'}^{q}(m+1)
\ee
where 
\be
 {\cal N}_{a,b}^{c}(m) =  \left \{ \begin{array}{l@{\enspace}l}
 1 & \mbox{ if } |a-b|+1\le c\le {\rm min}(a+b-1,2m-a-b-1) \\
  & \mbox{ and } a+b+c \mbox{ is odd}\\
 0& \mbox{ otherwise}
 \end{array}
 \right . 
\ee
The fusion ring contains two subrings generated by fields of the form $\phi_{1,s}$ and $\phi_{r,1}$ respectively. 
The two subrings intersect over a subring generated by the identity field and the operator $\phi_{1,m}\equiv \phi_{m-1,1}$. 
The bulk defects satisfy the same fusion rule. The defect 
\be 
{\cal S} \equiv {\cal D}_{1,m} \equiv {\cal D}_{m-1,1}
\ee
describes the spin reversal symmetry. It satisfies the group property 
\be
{\cal S}\circ {\cal S} = {\cal D}_{1,1} \, , 
\ee
fuses with the other defects according to 
\be
{\cal S} \circ {\cal D}_{r,s} = {\cal D}_{r,m+1-s} \equiv {\cal D}_{m-r,s} 
\ee
and acts on Cardy boundary conditions as 
\be
{\cal S}\cdot  (r,s) =(r,m+1-s) \equiv (m-r,s)\, . 
\ee
The spin reversal invariant Cardy boundary conditions are of the form $(\frac{m}{2},s)$, $s=1, \dots, \frac{m}{2}$ if $m$ is even and of the form 
$(r, \frac{m+1}{2})$, $r=1, \dots, \frac{m-1}{2}$ if $m$ is odd. For such boundary conditions we can introduce the  ${\cal S}$-charge for the boundary 
fields that according to (\ref{X1}) is given by 
\be
{\cal S} \psi^{[a,a]}_{i} = F_{aa}\left[ \begin{array}{cc} (m-1,1) &a\\
a& i
\end{array} \right] \psi^{[a,a]}_{i} = \pm  \psi^{[a,a]}_{i} 
\ee
where the boundary label $a$ is $(\frac{m}{2},s)$ or $(r, \frac{m+1}{2})$ depending on the parity of $m$ and $i\in a\times a$. 
This charge is equal to $\pm1$  due to the orthogonality relation (\ref{ort}) and the fusion rule $(m-1,1)\times a = a$. 

If we are perturbing an ${\cal S}$-invariant boundary condition by a charge 1 boundary field then, by virtue of the Graham-Watts 
theorem, we expect each end point of the flow to be ${\cal S}$-invariant. If we perturb by a charge -1 field then the end points 
of the flow are interchanged by the action of ${\cal S}$. For example in the tricritical Ising model, that corresponds to  $m=4$, 
we have two spin reversal invariant Cardy boundary conditions: $(2,2)$  and $(2,1)$. The latter boundary condition is stable while  the $(2,2)$ boundary condition, also known as the disordered boundary condition,   admits two relevant boundary fields: $\psi_{1,2}$ 
and $\psi_{1,3}$. The first field has the ${\cal S}$-charge -1 while the second one has charge 1.  
The  boundary RG flows in the tricritical Ising model that start from the elementary boundary conditions were described in \cite{Affleck}. Both $\psi_{1,2}$ 
and $\psi_{1,3}$  perturbations of the disordered boundary condition are integrable and their end points are given on the following diagrams:
\bea
&&  (2,1)  \stackrel{\psi_{1,3}}{\longleftarrow} (2,2)  \stackrel{-\psi_{1,3}}{\longrightarrow} (1,1)\oplus (3,1) \, , \nonumber \\ 
&&          (1,1)      \stackrel{\psi_{1,2}}{\longleftarrow} (2,2)  \stackrel{-\psi_{1,2}}{\longrightarrow} (3,1) \, .
\eea
It is straightforward to check that the endpoints satisfy the requirements for the action of ${\cal S}$.

Below we will be particularly interested in boundary flows triggered by perturbing the boundary condition $(r,s)$ by the boundary field 
$\psi_{1,3}^{[rs,rs]}$. For large values of $m$ these flows were studied in \cite{RRS} where the end points were identified using  the
$g$-theorem. The end points in the non-perturbative regime were found in \cite{GW} with the help of Graham-Watts theorem, which was put 
forward in that paper, and using the results of \cite{LSS}. The general rule for the end points of the $\psi_{1,3}$ flows that start from 
elementary boundary conditions can be summarised in the following two expressions 
\be \label{131}
(r,s) \longrightarrow \bigoplus_{i=1}^{{\rm min}(r,s,m-r,m-s)}(|r-s|+2i-1,1) \, , 
\ee
\be\label{132}
(r,s) \longrightarrow \bigoplus_{i=1}^{{\rm min}(r,s-1,m-r,m-s+1)}(|r-s+1|+2i-1,1) \, 
\ee
where one expression corresponds to a positive choice of the coupling and the other to the negative choice. To the best 
of our knowledge it has not been fixed in general which answer corresponds to which sign.  The  expressions (\ref{131}) 
and (\ref{132})  are interchanged under the action of the field identification  (\ref{Kac_identification}).

 Commutators of boundary fields with open defects can be computed using the general expression (\ref{A_shrink}). The fusion matrices 
 for the diagonal minimal models can be  calculated recursively following  \cite{Runkel} (see also \cite{Runkel_PhD} for a closed expression). 
 
\subsection{Tetracritical Ising model} 
The first example of a non-trivial open defect that is different from ${\cal S}$ and commutes with a relevant operator on an elementary boundary condition 
appears in the tetracritical Ising model that is the unitary minimal model with $m=5$. This model has 10 primary fields and 
thus the same number of topological defects and elementary conformal boundary conditions. We focus on the $\psi_{1,3}$ boundary 
field where we know the end points of the flows. All elementary boundary conditions have a $\psi_{1,3}$ boundary field except 
for the 4 boundary conditions of the form  $(r,1)$, $1\le r\le 4$. Table \ref{table-tetra} shows the open defects that have a topological junction 
with a given boundary condition and that commute or anti-commute  with $\psi_{1,3}$. 

\begin{table}[h!]
\begin{center}
\begin{tabular}{|c|c|c|}
\hline  
\rule{0pt}{4ex}  \rule[-3ex]{0pt}{0pt}  b.c.    &   defects commuting with $\psi_{1,3}$ &  defects anti-commuting with $\psi_{1,3}$\\
\hline 
\rule{0pt}{4ex}  \rule[-3ex]{0pt}{0pt} (1,3) & ${\cal D}_{1,1}^{[1,3]}$ & ${\cal S}^{[1,3]}$\\
\hline
\rule{0pt}{4ex}  \rule[-3ex]{0pt}{0pt} (3,3) &$ {\cal D}_{1,1}^{[3,3]}$, ${\cal D}^{[3,3]}_{3,1}$  & ${\cal S}^{[3,3]}$, ${\cal D}^{[3,3]}_{2,1}$\\
\hline
\rule{0pt}{4ex}  \rule[-3ex]{0pt}{0pt} (2,2) &${\cal D}_{1,1}^{[2,2]}$, ${\cal D}_{3,1}^{[2,2]}$   & none \\
\hline
\rule{0pt}{4ex}  \rule[-3ex]{0pt}{0pt} (3,2) &${\cal D}_{1,1}^{[3,2]}$, ${\cal D}_{3,1}^{[3,2]}$   & none \\
\hline
\end{tabular}
\caption{Open defects on boundary conditions in tetracritical Ising model}
\label{table-tetra}
\end{center}
\end{table}

We note that $(1,3)$ and $(3,3)$ boundary conditions are stable under fusion with ${\cal S}$ and thus are 
spin reversal symmetric while $(2,2)$ and $(3,2)$ form a doublet. The two boundary conditions omitted from the table: 
$(1,2)$, $(1,4)$, have no non-trivial defects commuting or anti-commuting with $\psi_{1,3}$. 

In view of the general discussion in section \ref{general_sec} for the $\psi_{1,3}$ flows that start from $(3,3)$, $(2,2)$ or $(3,2)$ 
boundary condition the end points must admit a topological junction with ${\cal D}_{3,1}$. Examining the fusion rules we 
find that this implies that they must contain 
one of the following 5 elementary boundary conditions: $(3,1)$, $(2,1)$, $(3,3)$, $(2,2)$, $(3,2)$.

Moreover, for the flows from $(3,3)$ and $(1,3)$ the end points are exchangeable by the fusion with ${\cal S}$. 
Also for the flows from $(3,3)$  there is a topological 
junction between the two end points and ${\cal D}_{2,1}$. These conditions become even more restrictive if we add constraints from the $g$-theorem.  The end points of the flows given by (\ref{131}), (\ref{132}) for the flows at hand are 
\bea
(2,1)\longleftarrow &(1,3)& \longrightarrow (3,1)\\
(4,1)\oplus(2,1) \longleftarrow &(3,3)&  \longrightarrow  (1,1)\oplus (3,1) \\
(1,1)\oplus(3,1) \longleftarrow &(2,2)&  \longrightarrow   (2,1)\\
 (4,1)\oplus (2,1) \longleftarrow &(3,2)&  \longrightarrow (3,1) \, .
\eea
We check that these flows satisfy all of the  constraints following from table \ref{table-tetra}. 

It is interesting to calculate the boundary fusion rings formed by the defects in table  \ref{table-tetra}. 
The $(2,2)$, $(3,3)$ and $(3,2)$  boundary conditions have  the open defect ring consisting of defects commuting with 
$\psi_{1,3}$ generated by ${\cal D}^{[a]}_{3,1}$ with a single relation 
given by 
\be \label{31rel}
{\cal D}^{[a]}_{3,1} \circ {\cal D}^{[a]}_{3,1} = f {\cal D}^{[a]}_{1 ,1} + (1-f){\cal D}^{[a]}_{3,1} \, , \quad f=\frac{1}{2}(\sqrt{5}-1) \, .
\ee
 In fact (\ref{31rel}) holds for any boundary condition $a$ admitting a topological junction with ${\cal D}_{3,1}$. This fact is a simple 
consequence of 
the bulk fusion rule and the general identities (\ref{gen_ids}). 
Thus, if 
the end point of a $\psi_{1,3}$ flow contains an elementary boundary condition admitting a topological junction with ${\cal D}_{3,1}$ 
then it will satisfy the same composition rule (\ref{31rel}) as in the UV boundary condition. 

The $(3,3)$ boundary condition has additional open defects that anti-commute with $\psi_{1,3}$ 
that satisfy the following relations under fusion
\be
{\cal D}_{2,1}^{[3,3]}\circ {\cal D}_{3,1}^{[3,3]} = f{\cal S}^{[3,3]} + (1-f){\cal D}_{2,1}^{[3,3]} \, , 
\ee
\be
{\cal D}_{2,1}^{[3,3]}\circ {\cal D}_{2,1}^{[3,3]} = f{\cal D}_{1,1}^{[3,3]} + (1-f) {\cal D}_{3,1}^{[3,3]} \, 
\ee
and ${\cal S}^{[3,3]}$ fuses with the other open defects according to the bulk fusion rule. 

The $(3,3)$ boundary condition also has a boundary $\psi_{2,1}$ field which is relevant. To the best of our knowledge these flows have not been investigated before and the end points have not been identified. We find that this perturbation commutes with ${\cal S}^{[3,3]}$ and 
anti-commutes with ${\cal D}_{1,3}^{[3,3]}$. The commutation with the spin reversal defect implies that 
each of the end points must be invariant under the spin reversal. Together with  the constraints from the 
$g$-theorem this gives us two possible 
infrared end points: $(1,3)$ and $(1,1)\oplus (1,5)$. The existence of a junction with ${\cal D}_{1,3}$  gives us two 
possible pairs of fixed points: either they are both $(1,3)$ or one of them is $(1,3)$ and the other is $(1,1)\oplus (1,5)$. 
Interestingly the condition on the $g$-factors being the same is satisfied for the second pair due to the identity 
$g_{1,3}=2g_{1,1} = 2g_{1,5}$. Moreover, we calculate the UV fusion of the anti-commuting defect  to be given by 
\be
{\cal D}^{[3,3]}_{1,3}\circ {\cal D}^{[3,3]}_{1,3} = \frac{1}{2} {\cal D}^{[3,3]}_{1,1} +\frac{1}{2}{\cal S}^{[3,3]} \, .
\ee
The same fusion rule must be satisfied by the ${\cal D}_{1,3}$ defect between the two infrared fixed points. 
It is straightforward to check that ${\cal D}_{1,3}^{[1,3]}$ satisfies the same rule. 
For the second pair we find that there is a unique combination 
\be
{\cal D}_{13}^{\rm IR} = \frac{1}{\sqrt{2}} ({\cal D}_{1,3}^{13,11} + {\cal D}_{1,3}^{13,15})
\ee
that satisfies\footnote{In checking these relations it is important to allow $a'$ and $b'$ on Figure \ref{open_fusion} 
each to take the values $(1,1)$ and $(1,5)$ independently of each other.} 
\bea
{\cal D}_{13}^{\rm IR} \circ {\cal D}_{13}^{{\rm IR}\dagger} &=& \frac{1}{2}({\cal D}_{1,1}^{[1,3]} + {\cal S}^{[1,3]})  \, , 
\nonumber \\
{\cal D}_{13}^{{\rm IR}\dagger} \circ {\cal D}_{13}^{\rm IR} &=& \frac{1}{2}[ ({\cal D}_{1,1}^{[1,1]} + {\cal D}_{1,1}^{[1,5]}) 
+ ({\cal S}^{[11,15]} + {\cal S}^{[15,11]})]
\eea
where ${\cal D}_{13}^{{\rm IR}\dagger }=({\cal D}_{1,3}^{11,13} + {\cal D}_{1,3}^{15,13})/\sqrt{2}$ is the conjugate defect. 
Thus, all constraints from the commuting and anti-commuting open defects are satisfied by each of the two pairs. 
We did check numerically\footnote{The author found an analytic argument based on RG interfaces that excludes the possibility that 
 both end points are $(1,3)$, but this is outside the scope of the present paper and will be reported elsewhere.}, using the truncated boundary conformal space approach of \cite{Watts_BTCSA}, that 
for positive $\lambda$ the flow at hand ends up at $(1,3)$ while for negative $\lambda$ it flows to $(1,1)\oplus (1,5)$.


\subsection{Pentacritical Ising model} \label{penta_sec}
Pentacritical Ising model corresponds to the minimal model with $m=6$. This model has 15 primary 
states and the same number of topological defects and conformal boundary conditions. Up to the action 
of the spin reversal generator we have 6 representatives of elementary boundary conditions admitting a boundary 
$\psi_{1,3}$ field: (1,2), (1,3), (2,2), (3,3), (2,3), (3,2). The boundary conditions $(3,3)$ and $(3,2)$ are spin-reversal 
invariant. The elementary boundary conditions that have non-trivial open defects commuting or anti commuting with $\psi_{1,3}$ 
are tabulated in table \ref{table-penta}. 
\begin{table}[h!]
\begin{center}
\begin{tabular}{|c|c|c|}
\hline  
\rule{0pt}{4ex}  \rule[-3ex]{0pt}{0pt}  b.c.    &   defects commuting with $\psi_{1,3}$ &  defects anti-commuting with $\psi_{1,3}$\\
\hline
\rule{0pt}{4ex}  \rule[-3ex]{0pt}{0pt} (3,3) &$ {\cal D}_{1,1}^{[3,3]}$, ${\cal D}^{[3,3]}_{3,1}$, ${\cal S}^{[3,3]}$  & none\\
\hline
\rule{0pt}{4ex}  \rule[-3ex]{0pt}{0pt} (2,2) &${\cal D}_{1,1}^{[2,2]}$, ${\cal D}_{3,1}^{[2,2]}$   & none \\
\hline 
\rule{0pt}{4ex}  \rule[-3ex]{0pt}{0pt} (2,3) & ${\cal D}_{1,1}^{[2,3]}$, ${\cal D}_{3,1}^{[2,3]}$  & none\\
\hline
\rule{0pt}{4ex}  \rule[-3ex]{0pt}{0pt} (3,2) &${\cal D}_{1,1}^{[3,2]}$, ${\cal D}_{3,1}^{[3,2]}$, ${\cal S}^{[3,2]}$    & none \\
\hline
\end{tabular}
\caption{Open defects on boundary conditions in pentacritical Ising model}
\label{table-penta}
\end{center}
\end{table}
We see that the end points of the flows that start with the 4 boundary conditions in table \ref{table-penta} (and their spin reverses) 
must admit a topological junction with ${\cal D}_{3,1}$. For  the spin reversal invariant boundary conditions: (3,3), (3,2), 
the end points must be also spin-reversal invariant. 
Noting that the $g$-factors satisfy $g_{3,3}>g_{3,2}>g_{3,1}$ we see that each end point 
of the $\psi_{1,3}$ flows from $(3,2)$  is either degenerate or is given by the $(3,1)$ boundary condition that is spin-reversal invariant. 

The expressions (\ref{131}), (\ref{132}) give  the flows 
\bea
(1,1)\oplus(3,1)\longleftarrow &(2,3)& \longrightarrow (2,1)\oplus(4,1)\\
(4,1)\oplus(2,1) \longleftarrow &(3,3)&  \longrightarrow  (1,1)\oplus (3,1)\oplus (5,1) \\
(1,1)\oplus(3,1) \longleftarrow &(2,2)&  \longrightarrow   (2,1)\\
 (4,1)\oplus (2,1) \longleftarrow &(3,2)&  \longrightarrow (3,1) \, .
\eea
It is straightforward to check that these flows satisfy the above constraints. 

It is interesting to take a look at the fusion rings of the open defects in table \ref{table-penta}.
Noting that  the bulk fusion rule 
\be
(3,1)\times (3,1) = (1,1) + (3,1) + (5,1) 
\ee
contains 3 terms, the boundary fusion rule (\ref{bfusion1}) now has room for different deformations. 
Indeed, we find 
\be \label{31fusion1}
{\cal D}^{[a]}_{3,1}\circ {\cal D}^{[a]}_{3,1} = \frac{1}{2}{\cal D}^{[a]}_{1,1} + \frac{1}{2}{\cal D}^{[a]}_{3,1}
\ee
for $a=(2,2), (4,4), (4,6), (2,6)$ and 
\be \label{31fusion2}
{\cal D}^{[b]}_{3,1}\circ {\cal D}^{[b]}_{3,1} = \frac{1}{2}{\cal D}^{[b]}_{1,1} + \frac{1}{2}{\cal S}^{[b]} 
\ee
for $b=(3,3), (3,2), (3,1)$. The ${\cal S}^{[b]}$ generator fuses according to the bulk fusion rule. 
It is interesting to note that the boundary fusion rule  (\ref{31fusion2}) means that $ {\cal D}^{[b]}_{3,1}$ is an open  duality defect in the sense of \cite{FFRS}, \cite{FFRS2},
 that is its fusion   with itself contains only group-like open defects. 
 
 We also have a boundary field $\psi_{1,2}$ present on the boundary condition $(3,3)$. The corresponding boundary flows are 
 believed to be integrable but, as in the case of $\psi_{2,1}$ perturbation in the Tetracritical model,  have not been 
 investigated before.  We find that ${\cal D}_{3,1}^{[3,3]}$ anti-commutes  and ${\cal S}^{[3,3]}$ commutes with $\psi_{1,2}$ 
 that makes this case quite similar to the case of $\psi_{2,1}$ perturbation considered at the end of the previous section. 
It is instructive to see how all of the consequences considered in  section \ref{general_sec} can be combined  with the constraints from the $g$-theorem to restrict the choices of the infrared fixed points. We can first list all spin reversal invariant boundary conditions  
with a $g$-factor lower than that of the UV value. This gives us two singlets: $(3,2)$, $(3,1)$; four doublets: 
$A=(3,1)\oplus (3,1)$, $B=(1,1)\oplus (5,1)$, $C=(5,5)\oplus(1,5)$, $D=(4,6)\oplus(2,6)$; one triplet: $(1,1)\oplus (5,1)\oplus(3,1)$; and one 
quadruplet: $(1,1)\oplus (1,1)\oplus (5,1)\oplus (5,1)$. The condition that the $g$-factors of both IR end points must be equal 
implies that either both end points are the same (and belong to the above list), or form one of the following 3 pairs: 
$$
(3,1) \mbox{  and } (1,1)\oplus(5,1) \, , \quad (3,1)\oplus (3,1) \mbox{  and } (1,1)\oplus(5,1)\oplus (3,1)\,  
$$
$$
\quad (3,1)\oplus (3,1) \mbox{  and } (1,1)\oplus (1,1)\oplus (5,1)\oplus (5,1)
$$
 that are permitted because of the identity: $g_{3,1}=2g_{1,1}=2g_{5,1}$. Adding the condition that there must be a topological 
 junction with ${\cal D}_{3,1}$ possible between the two end points discards two pairs with equal boundary conditions: $B$, $B$ and 
  $C$,  $C$. Finally, requiring that the junction with ${\cal D}_{3,1}$ should satisfy the fusion product given in (\ref{31fusion1}) we can 
  discard one more pair: $D$, $D$. This follows from checking that no combination of 3 available junctions: ${\cal D}_{3,1}^{[4,6]}$, 
  ${\cal D}_{3,1}^{[2,6]}$, ${\cal D}_{3,1}^{[46,26]}$ can be chosen to satisfy (\ref{31fusion1}). All of these constraints leave us in the 
  end with 7 distinct pairs of possible infrared fixed points. Thus, in this example we see that {\it each} of the constraints we derived in section \ref{general_sec} reduces the number of possibilities. We finish this example by reporting that truncated conformal space approach numerics gives the spectra that match with 
  the $(3,1)$ boundary condition for large positive $\lambda$ and that of $(1,1)\oplus (5,1)$ boundary condition for large negative $\lambda$. 

\section{Flows from direct sums of boundary conditions}
\label{sums_sec}
\setcounter{equation}{0}
So far we have discussed the implication of open defects commuting with the perturbation for the flows originating from an elementary boundary condition. 
This can be generalised to flows from direct sums of elementary boundary conditions triggered by boundary condition changing 
operators. Examples of such flows, including those triggered by $\psi_{1,3}$ operators,  were studied in \cite{Graham}. 
The commutator with an open defect has been calculated on Figure \ref{commutator_figure1}.

A new feature of direct sum boundary conditions  is that the perturbing field can be a linear combination of several 
components and a commuting (or anti-commuting) open defect can be a particular linear combination of defects with the same Virasoro 
label but linking different sets of elementary boundary conditions. One way to generate such linear combinations is by starting with 
a commuting open defect on an elementary boundary condition and fuse it with a closed defect. Suppose ${\cal D}^{[a]}_{d}$ 
commutes with $\psi_{i}^{[a,a]}$. We can fuse the junction with a closed string defect ${\cal D}_{s}$ on either  side of the junction. 
Such a fusion done 
on the left is illustrated on Figure \ref{left_fusion}.
\begin{figure}[H]
\begin{center}
\begin{tikzpicture}[scale=2,>=latex]
\draw [white,pattern=my north east lines,  line space=5pt, pattern color=black] (-1.1,0) rectangle (1.1,-0.1) ;
\draw[very thick, red] (0,1)--(0,0);
\draw[very thick, red,->] (0,0)--(0,0.6);
\draw (-0.7,0.09) node {$a$};
\draw (0.7,0.1) node {$a$};
\draw (0.12,0.6) node {$d$};
\draw[very thick] (-1.1,0)--(1.1,0);
\draw (0,0) node {$\bullet$} ;
\draw[very thick, blue] (-1.1,0.16)--(-0.3,0.16);
\draw[very thick, blue] (-0.3,0.16)--(-0.1,0.3);
\draw[very thick,blue] (-0.1,0.3)--(-0.1,1);
\draw[very thick,blue,->] (-0.1,0.6)--(-0.1,0.7);
\draw (-0.2,0.56) node {$s$};
\draw (1.3,0) node {$=$};
\draw [white,pattern=my north east lines,  line space=5pt, pattern color=black] (1.5,0) rectangle (3.7,-0.1) ;
\draw (3.2,0.12) node {$a$};
\draw[very thick] (1.5,0)--(3.7,0);
\draw[very thick, red] (2.6,0.4)--(2.6,0);
\draw[very thick, red,->] (2.6,0)--(2.6,0.3);
\draw[very thick, green] (2.6,0.4) -- (2.6,1);
\draw[very thick, green,->] (2.6,0.6) -- (2.6,0.8);
\draw (2.9,0.75) node {$s\times d$};
\draw[very thick, blue,->] (2.2,0)--(2.6,0.4);
\draw (2.6,0) node {$\bullet$} ;
\draw (2.6,0.4) node {$\bullet$} ;
\draw (2.2,0) node {$\bullet$} ;
\draw (2.3,0.3) node {$s$};
\draw (1.8,0.13) node {$s\times a$};
\end{tikzpicture}
\end{center}
\begin{center}
\begin{tikzpicture}[scale=2,>=latex]
\draw (-1,0) node {$=$};
\draw (0,0) node  {$\mathlarger{\sum\limits_{i \in s\times a}\sum\limits_{j \in s\times d} Y_{a,s,d}^{L;\,  i,j}}$};
\draw [white,pattern=my north east lines,  line space=5pt, pattern color=black] (1,0) rectangle (2.8,-0.1) ;
\draw[very thick] (1,0)--(2.8,0);
\draw[very thick, green] (1.9,0)--(1.9,0.9);
\draw[very thick, green,->] (1.9,0)--(1.9,0.6);
\draw (1.9,0) node {$\bullet$};
\draw (2.05,0.55) node {$j$};
\draw (1.4,0.14) node {$i$};
\draw (2.4,0.12) node {$a$};
\end{tikzpicture}
\end{center}
\caption{Fusion of a closed defect with a junction on the left}
\label{left_fusion}
\end{figure}
\noindent where the coefficients $Y_{a,s,d}^{L;\,  i,j}$ are easily computed using the results of \cite{S} (see Figure 8 of \cite{S} in particular). 
As the final configuration on Figure \ref{left_fusion} is only an intermediate result, we omit the explicit 
expression for  $Y_{a,s,d}^{L;\,  i,j}$. At this stage it is important for us to note that, as a consequence  of the commutation of 
${\cal D}^{[a]}_{d}$  with $\psi_{i}^{[a,a]}$, the open defect combination
\be \label{j_comb}
{\cal D}_{j}^{L} \equiv \sum_{i\in s\times a} Y_{a,s,d}^{L;\,  i,j} {\cal D}_{j}^{[i,a]}
\ee
satisfies 
\be \label{Djl}
{\cal D}_{j}^{L} \psi^{[a,a]}_{i} = {\cal D}_{s}(\psi_{i}^{[a,a]}) {\cal D}_{j}^{L} 
\ee
where 
\be \label{fused_field}
{\cal D}_{s}(\psi_{i}^{[a,a]}) = \sum\limits_{n,m\in s\times a} X_{i,nm}^{aa}\psi_{i}^{[n,m]}
\ee
is the action on  the boundary field $\psi_{i}^{[a,a]}$ of the  fusion of ${\cal D}_{s}$ with the boundary $a$. 
Note that (\ref{Djl}) is true for any fixed label $j$. Now, picking  a configuration given by   (\ref{j_comb}) 
with a fixed label $j$ we can further fuse it with the closed defect  ${\cal D}_{s}$ on the right side of the junctions. 
Using steps similar to those on Figure \ref{left_fusion} we arrive at the following open defect combinations 
\be \label{LRcomb}
{\cal D}_{l}^{j,LR} = \sum\limits_{n,m\in s\times a} Z^{asd}_{j(nml)}{\cal D}^{[n,m]}_{l}
\ee
that, due to the associativity of the fusion operations, commute with the fused boundary field (\ref{fused_field}) 
for each choice of the label $l\in s\times s\times d$ and $j\in s\times d$. The coefficients $Z^{asd}_{j(nml)}$ 
are calculated using the results of \cite{S}: 
\be
Z^{asd}_{j(nml)}= F_{aj}\left[ \begin{array}{cc} a&n\\
d& s
\end{array} \right]F_{al}\left[ \begin{array}{cc} m&n\\
s& j
\end{array} \right] \sqrt{\frac{F_{1n}\left[ \begin{array}{cc} s&a\\
s& a
\end{array} \right]F_{1m}\left[ \begin{array}{cc} s&a\\
s& a
\end{array} \right]}{F_{1n}\left[ \begin{array}{cc} l&m\\
l& m
\end{array} \right]}} {\cal N}^{asd}_{jl}
\ee
where 
\be
{\cal N}^{asd}_{jl} = \sqrt{\frac{F_{1a}\left[ \begin{array}{cc} d&a\\
d& a
\end{array} \right]}{F_{1j}\left[ \begin{array}{cc} s&d\\
s& d
\end{array} \right]F_{1l}\left[ \begin{array}{cc} s&j\\
s& j
\end{array} \right]}}
\ee
is an overall normalisation factor. 
Similarly, we can do the above fusion in the reversed order, that is first fusing with a closed defect on the right, 
singling out an elementary component labeled by $j$, then fusing it on the left and singling out open defects labeled by 
$l$. The resulting open defects are given by a combination 
\be\label{RLcomb}
{\cal D}_{l}^{j,RL} = \sum_{n,m} \tilde Z_{asd}^{j(nml)}{\cal D}^{[n,m]}_{l}
\ee 
where 
\be
\tilde Z_{asd}^{j(nml)} =  Z_{asd}^{j(mnl)} \sqrt{\frac{F_{1n}\left[ \begin{array}{cc} l&m\\
l& m
\end{array} \right]}{F_{1m}\left[ \begin{array}{cc} l&n\\
l& n
\end{array} \right]}} \, .
\ee
These open defects also commute with the fused boundary field ${\cal D}_{s}(\psi_{i}^{[a,a]})$.

We illustrate the constructions (\ref{fused_field}), (\ref{LRcomb}), (\ref{RLcomb}) on a couple of explicit examples. 
Consider  the Cardy boundary condition $(2,2)$ in the tetracritical Ising model. The open defect ${\cal D}^{[2,2]}_{31}$ 
commutes with  $\psi_{13}^{[22,22]}$. Fusing the boundary with a closed defect ${\cal D}_{1,2}$ 
we obtain the direct sum $(2,1)\oplus(2,3)$. Up to an overall factor  the boundary field   $\psi_{13}^{[22,22]}$ 
is mapped to the combination 
\be \label{Psi}
\Psi \equiv \tilde \psi_{13}^{[23,23]} - 2(\tilde \psi^{[23,21]} _{13}+ \tilde \psi^{[21,23]}_{13}) 
\ee
where we use the notation 
\be
\tilde \psi^{[a,b]}_{i} = \frac{1}{\alpha^{ab}_{i}}\psi^{[a,b]}_{i} 
\ee
and the boundary fields $\psi^{[a,b]}_{i} $ are normalised as in \cite{S}. 
Since $(1,2)\times (3,1) = (3,2)$ we have only one value $j=(3,2)$ in (\ref{LRcomb}), (\ref{RLcomb}).
Using (\ref{LRcomb}) we find the open defects combinations 
\be
{\cal D}_{3,3}^{[23,23]} - {\cal D}_{3,3}^{[23,21]} -\sqrt{2}{\cal D}_{3,3}^{[21,23]} \, , 
\ee
\be
{\cal D}_{3,1}^{[23,23]} - {\cal D}_{3,1}^{[21,21]}
\ee
each of which commutes with (\ref{Psi}) as can be checked directly. 
Using (\ref{RLcomb}) gives the same combinations. These combinations are  fixed by the commutation condition  up to an overall factor.

Our second example starts with the same triple: $a=(2,2)$, $d=(3,1)$, $\psi_{13}^{[22,22]}$, but this 
time we fuse it with ${\cal D}_{2,1}$. This gives the direct sum of boundary conditions: $(1,2)\oplus (3,2)$ with a boundary field 
\be
\Psi' \equiv 7 \tilde \psi^{[32,32]}_{1,3} - 9\tilde \psi^{[12,12]}_{1,3} \, .
\ee
(Again for brevity we dropped the overall normalisation factor.) 
We now have two choices for $j$   in (\ref{LRcomb}),  (\ref{RLcomb}): $j\in \{ (2,1), (4,1)\}$. This gives us 4 particular 
linear combinations of the following three elementary open defects: 
\be \label{elem_defs}
{\cal D}_{3,1}^{[32,32]}\, , \quad  {\cal D}_{3,1}^{[32,12]}\, , \quad {\cal D}_{3,1}^{[12,32]}\, .
\ee
 The coefficients of these linear combinations
are quite ugly so we do not present them here, but we checked that they span the linear subspace 
generated by the  elementary open defects listed in (\ref{elem_defs}). Indeed, a separate calculation shows  that each of 
the defects in (\ref{elem_defs}) commutes with $\Psi'$. 

  \section{Concluding remarks} \label{conc_sec}
  \setcounter{equation}{0}
Our considerations in section \ref{general_sec} did not depend on any particular choice of a rational CFT. Given 
an open topological defect on a conformal boundary  that either commutes or anti-commutes with 
a relevant boundary perturbation all the consequences for RG flows derived in that section would apply. 
By working out a number of explicit examples in the minimal models we showed that all of these constraints can be used to restrict
the possible infrared fixed points in RG flows, in particular in situations in which no other analytic arguments are known that would give the same restrictions. 

It would be interesting to generalise the calculations done in \cite{S} to a more general chiral algebra and to find 
other examples of applications of our general results to boundary RG flows in other models. More systematically, one can 
try to obtain some general results  towards classifying  all possible pairs - a relevant boundary operator plus a commuting or anti-commuting 
open topological defect, in given RCFTs or classes of RCFTs.  Certainly such situations, when such a pair exists, are special. 
As we discussed in section  \ref{general_sec},  boundary operators in such pairs generate a subalgebra under OPE. 
In the bulk CFT perturbations with this property the Hamiltonian is block diagonal that signals the presence of additional conserved charges. Moreover, 
like the $\Phi_{1,3}$, $\Phi_{1,2}$ and $\Phi_{2,1}$ bulk perturbations of minimal models  (see \cite{Zam_int}),   such 
perturbations are known to give integrable models. The integrability aspect of 
boundary perturbations is still comparatively less studied, particularly for the $\psi_{1,2}$ and $\psi_{2,1}$ perturbations. 
It would be interesting to investigate  possible connections between the presence of commuting or anti-commuting defects 
and integrability, perhaps one could try to exploit the link between defects and integrability established for bulk perturbations in 
\cite{Ingo_nonloc}. We hope to address these questions in future work.

\begin{center}
{\bf \large Acknowledgements} 
\end{center}
The author thanks M. Buican, I. Runkel and C. Schmidt-Colinet   for useful discussions and I. Runkel for comments on the draft. 

\appendix
\renewcommand{\theequation}{\Alph{section}.\arabic{equation}}
\setcounter{equation}{0}
\section{Some identities for the minimal model fusion matrices}
\be\label{A1}
F_{pq}\left[ \begin{array}{cc} a&b\\
c& d
\end{array} \right]= F_{pq}\left[ \begin{array}{cc} c&d\\
a& b
\end{array} \right] = F_{pq}\left[ \begin{array}{cc} b&a\\
d& c
\end{array} \right]
\ee
\be\label{A2}
F_{11}\left[ \begin{array}{cc} a&a\\
a& a
\end{array} \right] = \frac{S_{11}}{S_{1a}}
\ee
where $S_{ab}$ stand for the elements of the modular $S$-matrix. 
\be\label{A3}
F_{1a}\left[ \begin{array}{cc} b&c\\
b& c
\end{array} \right]  F_{a1}\left[ \begin{array}{cc} b&b\\
c& c
\end{array} \right]  =  \frac{S_{11}S_{1a}}{S_{1b}S_{1c}}
\ee
\be\label{A4}
F_{a1}\left[ \begin{array}{cc} b&b\\
c& c
\end{array} \right] =\frac{S_{1a}}{S_{1c}} 
F_{c1}\left[ \begin{array}{cc} a&a\\
b& b
\end{array} \right] 
\ee
\be\label{A5}
F_{e1}\left[ \begin{array}{cc} a&a\\
d& d
\end{array} \right] F_{fa}\left[ \begin{array}{cc} b&c\\
e& d
\end{array} \right] = F_{f1}\left[ \begin{array}{cc} c&c\\
d& d
\end{array} \right] F_{ec}\left[ \begin{array}{cc} a&b\\
d& f
\end{array} \right] 
\ee
\be \label{ort}
\sum_{s}F_{ps}\left[ \begin{array}{cc} b&c\\
a& d
\end{array} \right] F_{sr}\left[ \begin{array}{cc} c&d\\
b& a
\end{array} \right] =\delta_{pr}
\ee

We next show how one can use the above identities to establish the equivalence of the expression  for 
the coefficients $N_{cd}^{[a]\, e} $ that follows from (\ref{U_coefs}) and the expression presented in formula (\ref{bfusion2}). 
Formula (\ref{U_coefs})   gives 
\be \label{S_formula}
N_{cd}^{[a]\, e}= \frac{F_{1a}\left[ \begin{array}{cc} d&a\\
d& a
\end{array} \right]F_{1a}\left[ \begin{array}{cc} c&a\\
c& a
\end{array} \right] }{F_{1a}\left[ \begin{array}{cc} e&a\\
e& a
\end{array} \right]F_{1e}\left[ \begin{array}{cc} d&c\\
d& c
\end{array} \right]}F^2_{ae}\left[ \begin{array}{cc} d&c\\
a& a
\end{array} \right] \, .
\ee
To show that this equals the expression in (\ref{bfusion2}) we first use the identities 
\be
F_{ae}\left[ \begin{array}{cc} d&c\\
a& a
\end{array} \right] = \frac{F_{da}\left[ \begin{array}{cc} e&a\\
c& a
\end{array} \right]F_{a1}\left[ \begin{array}{cc} a&a\\
c& c
\end{array} \right]}{F_{d1}\left[ \begin{array}{cc} e&e\\
c& c
\end{array} \right]} \, ,
\ee
\be
F_{ae}\left[ \begin{array}{cc} d&c\\
a& a
\end{array} \right] = \frac{F_{ca}\left[ \begin{array}{cc} e&a\\
d& a
\end{array} \right]F_{a1}\left[ \begin{array}{cc} a&a\\
d& d
\end{array} \right]}{F_{c1}\left[ \begin{array}{cc} e&e\\
d& d
\end{array} \right]}
\ee
that are particular instances of  (\ref{A5}). Substituting each of these identities into (\ref{S_formula}) we obtain 
\be \label{UtildeU}
N_{cd}^{[a]\, e}= \frac{F_{da}\left[ \begin{array}{cc} e&a\\
c& a
\end{array} \right]F_{ca}\left[ \begin{array}{cc} e&a\\
d& a
\end{array} \right]}{F_{1a}\left[ \begin{array}{cc} e&a\\
e& a
\end{array} \right]} \tilde N_{cd}^{[a]\, e}
\ee
where
\be \label{tildeU}
\tilde N_{cd}^{[a]\, e} = \frac{F_{1a}\left[ \begin{array}{cc} d&a\\
d& a
\end{array} \right]F_{a1}\left[ \begin{array}{cc} a&a\\
d& d
\end{array} \right]    F_{1a}\left[ \begin{array}{cc} c&a\\
c& a
\end{array} \right]F_{a1}\left[ \begin{array}{cc} a&a\\
c& c
\end{array} \right]}{F_{1e}\left[ \begin{array}{cc} d&c\\
d& c
\end{array} \right]F_{d1}\left[ \begin{array}{cc} e&e\\
c& c
\end{array} \right]F_{c1}\left[ \begin{array}{cc} e&e\\
d& d
\end{array} \right]}
\ee
Using (\ref{A3}) in the numerator of (\ref{tildeU}) we rewrite the last expression as 
\be \label{tildeU2}
\tilde N_{cd}^{[a]\, e} = \frac{S_{11}^2}{S_{1d}S_{1c}F_{1e}\left[ \begin{array}{cc} d&c\\
d& c
\end{array} \right]F_{d1}\left[ \begin{array}{cc} e&e\\
c& c
\end{array} \right]F_{c1}\left[ \begin{array}{cc} e&e\\
d& d
\end{array} \right]}
\ee
Finally we use the two identities 
\be \label{2ids}
F_{d1}\left[ \begin{array}{cc} e&e\\
c& c
\end{array} \right] = \frac{S_{11}}{S_{1c}F_{1e}\left[ \begin{array}{cc} d&c\\
d& c
\end{array} \right]} \, , \qquad 
F_{c1}\left[ \begin{array}{cc} e&e\\
d& d
\end{array} \right] = \frac{S_{11}}{S_{1d}F_{1e}\left[ \begin{array}{cc} d&c\\
d& c
\end{array} \right]}
\ee
 that follow from a combination of (\ref{A3}) and (\ref{A4}). Substituting (\ref{2ids}) into (\ref{tildeU2}) 
 we obtain 
 \be
 \tilde N_{cd}^{[a]\, e} = F_{1e}\left[ \begin{array}{cc} d&c\\
d& c
\end{array} \right]
 \ee
 that being combined with (\ref{UtildeU}) gives (\ref{bfusion2}).
 
 Alternatively, formula (\ref{bfusion2}) can be obtained independently by using a sequence of moves on 
 the topological defects involved, different from the ones used in \cite{S}.

\end{document}